\documentclass{aa}  

\usepackage{graphicx}
\usepackage{txfonts}
\usepackage[version=4]{mhchem}
\usepackage{hyperref}

\usepackage{xcolor}
\begin{document}

\title{Dust enrichment from core-collapse supernovae and\\ extinction curves in the high-redshift universe}

\author{
        Koki Otaki
        \inst{1}
    \and
        Raffaella Schneider
        \inst{2,3,4,5}
    \and
        Simone Bianchi
        \inst{6}
    \and\\
        Joris Witstok
        \inst{7,8}
    \and
        Luca Graziani
        \inst{2,3,4}
    \and
        Marco Limongi
        \inst{3,9,10}
    \and
        Roberto Maiolino
        \inst{11,12,13}
      }
\institute{
        Information Technology Center, The University of Tokyo, 6-2-3 Kashiwanoha, Kashiwa, Chiba 277-0882, Japan\\
          \email{koki.otaki@cc.u-tokyo.ac.jp}
    \and
        Dipartimento di Fisica, ``Sapienza'' Universit\`{a} di Roma, Piazzale Aldo Moro 5, I-00185 Roma, Italy
    \and
        INAF/Osservatorio Astronomico di Roma, Via Frascati 33, I-00040 Monte Porzio Catone, Italy
    \and
        INFN, Sezione Roma I, Dipartimento di Fisica, ``Sapienza'' Università di Roma, Piazzale Aldo Moro 2, I-00185, Roma, Italy
    \and
        Sapienza School for Advanced Studies, Viale Regina Elena 291, I-00161 Roma, Italy
    \and
        INAF/Osservatorio Astrofisico di Arcetri, Largo E. Fermi 5, I-50125 Firenze, Italy
    \and
        Cosmic Dawn Center (DAWN), Copenhagen, Denmark
    \and
        Niels Bohr Institute, University of Copenhagen, Jagtvej 128, DK-2200, Copenhagen, Denmark
    \and
        Kavli Institute for the Physics and Mathematics of the Universe (WPI), The University of Tokyo Institutes for Advanced Study, The University of Tokyo, Kashiwa, Chiba 277-8583, Japan
    \and
        INFN, Sezione di Perugia, via A. Pascoli s/n, I-06125 Perugia, Italy
    \and
        Kavli Institute for Cosmology, University of Cambridge, Madingley Road, Cambridge CB3 0HA, UK
    \and
        Cavendish Laboratory, University of Cambridge, 19 JJ Thomson Avenue, Cambridge CB3 0HE, UK
    \and
        Department of Physics and Astronomy, University College London, Gower Street, London WC1E 6BT, UK
}

\date{}

\abstract
{
Recent JWST observations have revealed that some galaxies at $z\gtrsim7$ generally exhibit relatively flat ultraviolet (UV) attenuation curves and a weak UV bump. These features suggest that the first dust grains formed rapidly, possibly originating from core-collapse supernovae (SNe). We investigate the time evolution of grain size distributions and extinction curves in the early phase of dust enrichment for different parameters of progenitor stars, rotation velocities, metallicity, and interstellar medium densities, including the effect of the reverse shock.
We model a single starburst system assuming an initial mass function. 
Extinction curves are calculated from the grain size distribution for each dust species. 
The total dust-to-stellar mass ratio at $30\,\mathrm{Myr}$ is $M_\mathrm{dust}/M_\star\sim10^{-3}$ before the passage of the reverse shock, but we find it to be at most $M_\mathrm{dust}/M_\star\sim10^{-5}$ due to the destruction effect of the reverse shock.
This effect destroys grains smaller than $\sim10\,\mathrm{nm}$ and makes amorphous carbon the dominant species, resulting in a flatter extinction curve with a wide bump at $2500\,\text{\AA}$ compared to the no-reverse shock models. We find that our models are consistent with the observed attenuation curve and emissivity of high-redshift galaxies and show that the reverse shock processing significantly affects dust enrichment and grain properties such as extinction curves and emissivity in supernova yields for high-redshift galaxies.
}


\maketitle
\nolinenumbers


\section{Introduction}
\label{sec:intro}

Interstellar dust plays important roles in galaxy evolution and in determining the ultraviolet (UV) and optical properties of galaxies, as it can absorb UV photons, depending on the dust column density and geometry, and re-emits them in the infrared (IR).
The surface of dust grains serves as a site for $\ce{H2}$ formation in the interstellar medium (ISM) \citep{Stahler2005}.
In regions dominated by stellar UV photons, dust acts as a heating source, whereas in dense regions, infrared emission from dust contributes to the cooling of the ISM \citep{draine2004}.
Dust-induced gas fragmentation in low-metallicity regions plays a crucial role in determining the nature of stellar populations \citep{schneider2003, schneider2006, schneider12b, schneider12a, omukai2005, marassi2014, chiaki2014, chiaki2015}.
From a dynamical perspective, dust grains also transfer momentum to the gas through radiation pressure from stars.

The effects of dust absorption and scattering for the UV and optical photons are reflected in the observed dust extinction curve, which incorporates physical properties of dust, such as the grain size and composition \citep{draine2003}. 
Empirically, for the Local Group, the extinction properties have been well-studied in the Milky Way (MW; \citealt{cardelli1989, fitzpatrick1990, pei1992, valencic2004, fitzpatrick2007, gordon2023}), the Large Magellanic Cloud (LMC; \citealt{clayton1985, fitzpatrick1985, misselt1999, gordon2003}), the Small Magellanic Cloud (SMC; \citealt{gordon1998, gordon2003, maiz2012, gordon2024}) and M31 \citep{bianchi1996, clayton2015, wang2022, clayton2025} through observations combined with theoretical models.
The extinction curve in local galaxies generally increases with the inverse of wavelength. The MW, LMC, and M31 exhibit similar curves characterized by a prominent bump at $2175\,\text{\AA}$, whereas the curve of the SMC is noticeably steeper than those of the other local galaxies.
This UV bump is attributed to polycyclic aromatic hydrocarbons (PAHs), although other forms of carbonaceous grains may also contribute, and its strength is influenced by the metallicity, age, and stellar mass of the galaxy. In particular, the extinction curve of M31 shows a weaker bump and a steeper far-UV slope compared to that of the MW, suggesting a trend within the Local Group in which the UV slope of the extinction curve increases as the bump strength decreases \citep{gordon2024, clayton2025}.

Beyond the Local Group, the attenuation properties of star-forming galaxies are often characterized by the empirical Calzetti attenuation curve \citep{calzetti2000}, derived from observations of local starburst galaxies. Unlike the extinction curves of the MW, LMC, and M31, the Calzetti attenuation curve lacks the prominent $2175\,\text{\AA}$ UV bump and exhibits a shallower rise toward shorter wavelengths compared to the SMC, indicating weaker UV attenuation per unit reddening. This difference arises because the Calzetti curve represents the attenuation of integrated galaxy light - accounting for both dust absorption and scattering, as well as the complex geometry between stars and dust - rather than the pure extinction along a single line of sight. The Calzetti law is widely applied to correct the spectral energy distributions of star-forming galaxies, particularly at high redshift. 

The advent of JWST has enabled substantial progress in constraining the attenuation curves of high-redshift galaxies. These observations have revealed UV bumps at very early cosmic epochs, suggesting the rapid formation of carbonaceous dust grains and PAHs in young galaxies \citep{witstok2023b, markov2023, yang2023, sanders2025, Ormerod2025, lin2025}, and have indicated that the
UV-optical slopes of attenuation curves at high redshift are, on average, relatively shallow
\citep{Langeroodi2024, Markov2025, Fisher2025, McKinney2025, shivaei2025,Rodighiero2026}. 

Interestingly, the highest-redshift galaxies at $z = 7 - 9$ exhibit slopes that are flatter than the Calzetti curve, with important implications for their UV obscuration and IR emission \citep{shivaei2025}.
At the same time, other JWST observations suggest that small dust grains may already be abundant in the circumgalactic medium of massive galaxies at high redshift, leading to steep extinction curves comparable to or even steeper than that of the SMC \citep{Sun2026}. These seemingly diverse observational results highlight the complex interplay between dust production, processing, and spatial distribution in the early Universe.

While it is tempting to interpret the JWST findings as evidence for a changing nature of the dominant dust sources, it is essential to recognize that the geometry of the dust distribution relative to the stars plays a crucial role in determining the observed dust attenuation. Distinct geometrical configurations can yield similar attenuation curves even when the underlying extinction curves differ substantially, and conversely, similar dust properties may produce different attenuation curves depending on the geometry \citep{Gallerani2010,Lin2021}.

In the theoretical context of dust formation in the early Universe, supernova (SN) explosions are recognized as major contributors, ejecting newly formed grains into the surrounding ISM on timescales of approximately 
$\leq 30 \,\mathrm{Myr}$ (see \citealt{schneider2024} for a recent review). The grains formed in SN ejecta are generally believed to have relatively large radii ($\gtrsim0.1\,\mathrm{\mu m}$), resulting in flatter extinction curves \citep{maiolino2004, hirashita2005, bianchi2007, hirashita2008}. Over time, grain reprocessing in the ISM leads to a steepening of the UV–optical slope, while additional carbon dust production from asymptotic giant branch (AGB) stars may contribute to the emergence of the UV bump (see, e.g., \citealt{Hirashita2019, Hirashita2020}). 

Hydrodynamical simulations of galaxy formation that incorporate models for the evolution of the dust mass indicate that stellar dust production dominates dust enrichment at the highest redshifts ($z > 10 \text{-} 11$, \citealt{Narayanan2025}) and galaxies with the lowest stellar masses ($M_{\star} < 10^8\text{-}10^9 \, M_\odot$, \citealt{graziani2020, dicesare2023}). As a consequence, in galaxies where dust production is primarily driven by SNe, the grain size distribution tends to be skewed toward larger grains \citep{Dubois2024, Narayanan2025}, resulting in UV-optical attenuation curves that are shallower than the SMC and the Calzetti curve, and in good agreement with JWST observations \citep{shivaei2025}. 

It is therefore very important to characterize the extinction curves expected when dust production occurs on short timescales ($\lesssim 30\,\mathrm{Myr}$) and driven by SNe. The dust yields from SNe have been investigated in numerous studies using either classical or kinetic nucleation theory. However, not all of the dust grains formed in SN ejecta survive to be injected into the ISM; the reverse shock, generated by the interaction between the forward shock and the ambient medium, can efficiently destroy a significant fraction of the newly condensed dust (see recent reviews by \citealt{micelotta2018, schneider2024}).
 
Recently, \citet{otaki2026} computed the effective SN dust yields from nonrotating and rotating progenitor stars over a broad range of masses ($13\, M_\odot\leq m_\star \leq 120\, M_\odot$) and metallicity ($[\ce{Fe/H}]=0,\,-1,\,-2,$ and $-3$), incorporating the destructive effects of the reverse shock. Leveraging these results, in the present study, we model the expected time evolution of the grain size distribution and extinction curve in galaxies dominated by young stellar populations (ages $< 30\,\mathrm{Myr}$) representative of some of the high-redshift galaxies observed with JWST.

The paper is organized as follows. In Section~\ref{sec:grid}, we briefly describe the grid of effective SN dust yields computed by \citet{otaki2026}. In Section~\ref{sec:model}, we outline our model for the time evolution of dust masses, grain size distributions, and extinction properties of single stellar populations. Section~\ref{sec:result} presents the main results, while Section~\ref{sec:discuss} discusses our findings in the context of previous studies and recent JWST observations. Finally, Section~\ref{sec:conclusions} summarizes our main conclusions.

\section{Grid of effective dust yields}
\label{sec:grid}

Here we briefly describe the main properties of the grid of effective SN yields used in the present work, based on the recent study by \citet{otaki2026}. We refer the interested reader to the original paper for more details.

The dust yields from SNe prior to the passage of the reverse shock were previously provided by \citet{marassi2019}, who applied Classical Nucleation Theory \citep{bianchi2007, marassi2014, marassi2015} to a grid of SN models computed by \citet{limongi2018}.

The grid of SN models spans a range of initial progenitor masses ($13\, M_\odot\leq m_\star \leq 120\, M_\odot$), initial rotation velocities ($v=0$ and $300\,\mathrm{km\,s^{-1}}$), and initial metallicity ($[\ce{Fe/H}]=0,\,-1,\,-2,$ and $-3$), assuming a fixed explosion of $1.2\times10^{51}\,\mathrm{erg}$ for all the models. The pre-SN evolution is simulated using the \texttt{FRANEC} code \citep{chieffi2013, limongi2018}.

Before the passage of the reverse shock, some massive progenitors do not produce dust because their ejecta either lack sufficient material to form grains (failed SNe) or possess helium cores large enough to trigger pair production, leading to pulsational pair-instabilities, whose evolution the \texttt{FRANEC} code cannot properly track. This effectively limits the range of progenitor masses contributing to dust production in our models. Compared to nonrotating progenitors, rotation-induced mixing increases mass loss during the pre-SN evolution and results in more metal-enriched ejecta, leading to more efficient dust formation.

For each SN model, the mass, composition, and size distributions are computed for amorphous carbon (\ce{AC}), iron (\ce{Fe}), corundum (\ce{Al2O3}), magnetite (\ce{Fe3O4}), enstatite (\ce{MgSiO3}), forsterite (\ce{Mg2SiO4}), and quartz (\ce{SiO2}).

The destructive effect of the reverse shock is calculated using the \texttt{GRASHrev} code  \citep{bocchio2016}, assuming that the SN explosions occur in a uniform ISM with densities $n_\mathrm{ISM}=0.05,\,0.5,$ and $5\,\mathrm{cm^{-3}}$. 

Before encountering the shock, the dust and gas in the ejecta are coupled, expanding homologously with the temperature decreasing adiabatically. 
However, the gas and dust respond differently and decouple when they are hit by the reverse shock.
Dust reprocessing is considered through both thermal and non-thermal sputtering. The resulting dust yields after the passage of the reverse shock depend on the grain size distribution prior to the shock (see \citealt{otaki2026} for a detailed discussion).

Before the passage of the reverse shock, the most efficient dust producers in the whole model grid are the solar-metallicity nonrotating model with $25\, M_\odot$ and the solar-metallicity rotating model with $15\, M_\odot$, producing up to $\sim 1.2\, M_\odot$ and $\sim2.2\, M_\odot$ of dust, respectively. 
However, the destructive effects of the reverse shock significantly reduce the dust masses. For instance, among all progenitors, the largest surviving dust masses after the passage of the reverse shock are found for the most massive stars. In particular, $120\, M_\odot$ progenitors with the same metallicity produce only $0.02\, M_\odot$ and $0.03\, M_\odot$ of dust in the nonrotating and rotating models, respectively, when embedded in an ISM with a density of $n = 0.5\,\mathrm{cm}^{-3}$. The surviving dust mass tends to decrease with increasing
ISM density, and larger grains $(\gtrsim10\,\mathrm{nm})$ are more resistant to the destructive effects of the reverse shock than smaller grains. Because of this, grains formed in SN ejecta from less massive stellar progenitors - which have generally smaller sizes - tend to be more heavily destroyed than those from more massive ones, regardless of their initial metallicity or rotation velocity.  

For each SN model, the surviving fraction of the initial dust mass after the reverse shock ranges from 0 to a maximum of 5\%. The reverse shock alters both the grain size distribution and composition. For a given progenitor mass, the average grain size tends to decrease with increasing ISM density. Although small grains are destroyed, sputtering also affects larger grains and modifies the grain size distribution. Since, in most models, AC grains are generally the largest prior to the passage of the reverse shock, they are more resistant to destruction and thus dominate the surviving dust mass. In contrast, small grains composed of other species are almost entirely destroyed by the reverse shock.

\section{Modeling dust properties}
\label{sec:model}


We model a single stellar population formed in an instantaneous burst at $t=0$, adopting an initial mass function (IMF) to calculate the time evolution of the dust masses, grain size distributions, and extinction curves. We consider three forms of the IMF, depending on the metallicity and redshift of the forming stars.

The Kroupa IMF \citep{Kroupa2001} is widely used for present-day stellar populations; however, in low-metallicity environments, metal-poor stars are believed to form according to a more top-heavy IMF due to the reduced cooling efficiency (see, e.g., \citealt{chon2021,chon2022}). For this reason, we adopt the Kroupa IMF for stellar populations with relatively high metallicity ($[\ce{Fe/H}] = 0, -1,$ and $-2$):
\begin{gather}
    \phi_\text{Kroupa}(m) \propto
    \begin{cases}
        m^{-0.3},& 0.01\,M_\odot \leq m < 0.08\,M_\odot,\\
        m^{-1.3},& 0.08\,M_\odot \leq m < 0.5\,M_\odot,\\
        m^{-2.3},& 0.5\,M_\odot \leq m \leq 120\,M_\odot,
\end{cases}
\end{gather}
and a log-flat IMF for the stellar population with the lowest metallicity ($[\ce{Fe/H}] = -3$):
\begin{gather}
    \phi_\text{log-flat}(m) \propto m^{-1},
\end{gather}
with the same lower and upper mass limits as the Kroupa IMF.
Moreover, the increasing temperature of the Cosmic Microwave Background (CMB) shifts the characteristic stellar mass toward more massive stars, introducing a redshift dependence of the stellar IMF (see, e.g., \citealt{schneider2010} and \citealt{chon2022}).
To model stellar populations in high-redshift galaxies, we therefore also consider the IMF proposed by \citet{chon2022}, which is characterized by a metallicity- and redshift-dependent mass fraction in the log-flat massive stellar component:
\begin{equation}
    f_\text{massive}=1.07\times (1-2^x) + 0.04\times 2.67^x \times z,
\end{equation}
where $x=1+\log Z/Z_\odot$.
We define the Chon IMF as a combination of the Kroupa IMF for stars lower than the critical mass $m_\text{crit}$ and the log-flat component for larger stellar mass:
\begin{gather}
    \phi_\text{Chon}(m) \propto 
    \begin{cases}
    m^{-0.3} ,& 0.01\,M_\odot \leq m < 0.08\,M_\odot,\\
    m^{-1.3} ,& 0.08\,M_\odot \leq m < 0.5\,M_\odot,\\
    m^{-2.3} ,& 0.5\,M_\odot \leq m < m_\text{crit},\\
    m^{-1},& m_\text{crit} \leq m \leq 120\,M_\odot,
    \end{cases}
\end{gather}
where we note that the upper mass limit of $120\,M_\odot$ is set by the available grid of SN yields. In this grid, some of the most massive progenitors do not contribute to dust production since they either fail to explode as SNe or undergo pulsational pair-instability, thereby limiting the mass range of dust-producing stars.
The IMF is normalized as $\int_{0.01 M_\odot}^{120 M_\odot} m\phi_\text{Chon} \, \mathrm{d}m= 1$, and $f_\text{massive}$ is computed as
\begin{gather}
    f_\text{massive} = \int_{m_\text{crit}}^{120\,M_\odot} m\phi_\text{Chon} \,\mathrm{d}m.
\end{gather}
Therefore, the critical mass is derived from a given value of $f_\text{massive}$, together with the normalization and junction conditions of the IMF.
\begin{figure}
    \centering
    \includegraphics[width=0.9\linewidth]{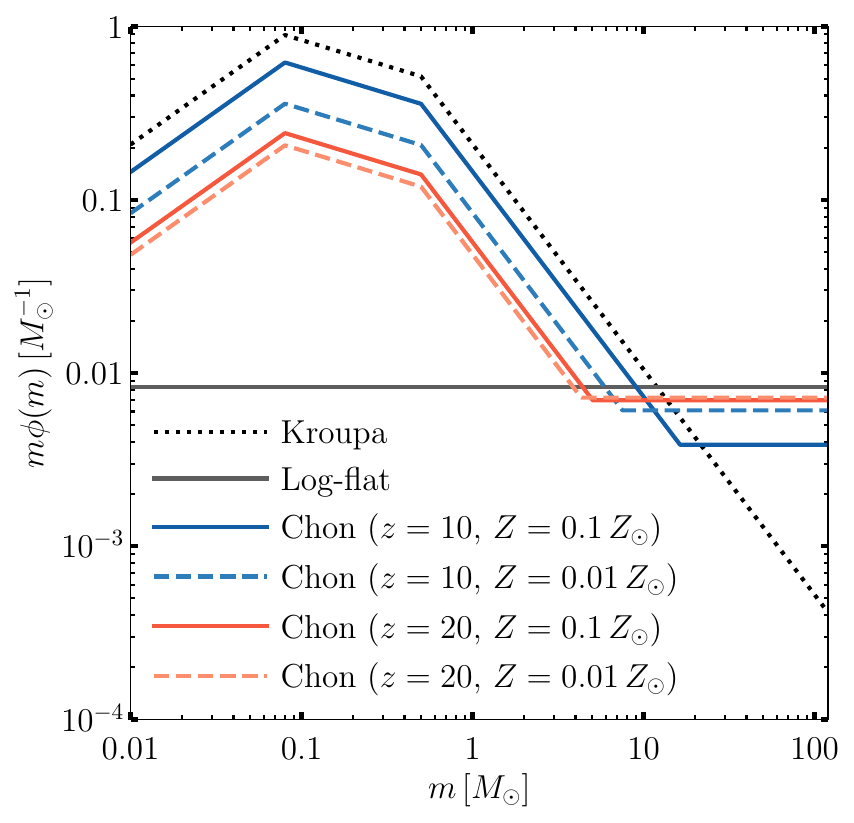}
    \caption{Initial mass functions adopted in this study. The black solid and dotted lines are the log-flat and the Kroupa IMFs. The red and blue lines represent the redshift- and metallicity-dependent Chon IMF with $(z,\,Z) = (10,\,0.1\, Z_\odot)$ (blue solid), $(10,\,0.01\, Z_\odot)$ (blue dashed), $(20,\,0.1\, Z_\odot)$ (red solid), and $(20,\,0.01\, Z_\odot)$ (red dashed), respectively. Each IMF is normalized so as $\int_{0.01\,M_\odot}^{120\,M_\odot}m\phi\,\mathrm{d}m=1$.}
    \label{fig:IMF}
\end{figure}

We show the IMFs in Figure~\ref{fig:IMF}.
The black dotted and solid lines represent the Kroupa and log-flat IMFs, respectively.
The Chon IMF is shown for different combinations of redshift $z$ and metallicity $Z$. The blue and red lines correspond to $z=10$ and $20$, and solid and dashed lines indicate $Z=0.1\, Z_\odot$ and $0.01\, Z_\odot$, respectively. Compared to the Kroupa IMF, the Chon IMF yields a higher mass fraction in massive stars. The critical mass for the Chon IMF increases with decreasing redshift and increasing metallicity, with a higher critical mass indicating a lower mass fraction of massive stars.

We define the number of massive stars that can potentially explode as supernovae\footnote{Depending on the initial stellar metallicity and rotation rate, the most massive progenitors of the grid are found to fail the explosion and to collapse to a black hole (see \citealt{limongi2018} for a thorough discussion).} per unit stellar mass $M_\star$ as:
\begin{gather}
f_\mathrm{SN} = \frac{N_\mathrm{SN}}{M_\star}=\int_{13\,M_\odot}^{120\,M_\odot}\phi(m)\,\mathrm{d}m.
\end{gather} 
For $(z,\, Z) = (10,\,0.1\, Z_\odot),\,(10,\,0.01\, Z_\odot),\,(20,\,0.1\, Z_\odot)$ and $(20,\,0.01\, Z_\odot)$, we find that $m_\mathrm{crit}=16,\,7.5,\, 5.0,$ and $4.3\, M_\odot$, and $f_\mathrm{SN} = 8.7\times10^{-3}\,{M_\odot^{-1}},\,1.4\times10^{-2}\,{M_\odot^{-1}},\,1.5\times10^{-2}\,{M_\odot^{-1}}$ and $1.6\times10^{-2}\,{M_\odot^{-1}}$, respectively. 
In contrast, $f_\mathrm{SN}=5.4\times10^{-3}\,{M_\odot^{-1}}$ for the Kroupa IMF. As a result, the supernova contribution to metal and dust enrichment in galaxies at $z \geq 10$ is larger when we adopt the Chon IMF to describe their stellar populations.
In this paper, we assume solar-scaled abundances $\log Z/Z_\odot\sim[\ce{Fe/H}]$.

For each given star, the dust mass released in grains with sizes in the interval [$a_j$, $a_j+\Delta a_j$] is calculated as:
\begin{gather}
    m_{\mathrm{dust},\,j} = \sum_{i\in \mathrm{species}} \int_{a_j}^{a_j+\Delta a_j}\left(\frac{\mathrm{d}n}{\mathrm{d}a}\right)_i \frac43\pi a^3\rho_i\,\mathrm{d}a,
\end{gather}
where $(\mathrm{d}n/\mathrm{d}a)_i$ is the dust size distribution of the $i$-th grain species, with $a_\mathrm{min}\leq a \leq a_\mathrm{max}$, $\Delta a_j$ is the size of the $j$-th bin, and $\rho_i$ is the grain density of the $i$-th grain species used in \citet{otaki2026} such as $\rho_{\ce{Al2O3}}=3.83\,\mathrm{g\,cm^{-3}}$, $\rho_{\ce{Fe}}=7.9\,\mathrm{g\,cm^{-3}}$, $\rho_{\ce{Fe3O4}}=5.12\,\mathrm{g\,cm^{-3}}$, $\rho_{\ce{MgSiO3}}=3.18\,\mathrm{g\,cm^{-3}}$, $\rho_{\ce{Mg2SiO4}}=3.32\,\mathrm{g\,cm^{-3}}$, $\rho_{\ce{AC}}=2.2\,\mathrm{g\,cm^{-3}}$, and $\rho_{\ce{SiO2}}=2.65\,\mathrm{g\,cm^{-3}}$.
Hence, the total dust mass released by the star is defined as $m_{\mathrm{dust}}=\sum_j m_{\mathrm{dust},\,j}$.

In this paper, we adopt SN dust yields and grain size distributions calculated by \citet{otaki2026} for each combination of stellar rotation, metallicity, and uniform ISM density. The stellar lifetime $\tau$ for each progenitor is taken from \citet{limongi2018} and depends on the stellar mass, rotation velocity, and metallicity. In general, the lifetimes of rotating stars are longer than those of nonrotating stars at fixed mass.

For a given star formation history, $\mathrm{SFR(t)}$, and a given IMF, the mass of dust released in grains with sizes in the interval [$a_j$, $a_j+\Delta a_j$] can be computed as:
\begin{gather}
    M_{\mathrm{dust},\,j}(t) = \int_0^t dt^\prime\int_{m_{\tau}(t^\prime)}^{120\,M_\odot}m_{\mathrm{dust},\,j}(m)\phi(m) \mathrm{SFR}(t^\prime -\tau(m))\,\mathrm{d}m,
    \label{eq:cosmicyield}
\end{gather}
where $m_\tau$ is the minimum mass of the star whose lifetime corresponds to the time $t'$.
Assuming that all stars form in a single burst at $t = 0$, the star formation rate becomes $\mathrm{SFR}(t)=M_\star \delta(t)$, thus Eq. \ref{eq:cosmicyield} reads: 
\begin{gather}
    M_{\mathrm{dust},\,j}(t) = M_\star \int_0^t dt^\prime\int_{m_{\tau}(t^\prime)}^{120\,M_\odot}m_{\mathrm{dust},\,j}(m)\phi(m) \delta(t^\prime -\tau(m))\,\mathrm{d}m,
\end{gather}
and the total dust mass expected at time $t$ can be computed as:
\begin{gather}
    M_\mathrm{dust}(t) = \sum_j M_{\mathrm{dust},\,j}(t).
\end{gather}

To compute the extinction curve, we use the \texttt{DustEM} code \citep{compiegne2011}. 
The absorption $Q_{\mathrm{abs},\,i}$ and scattering $Q_{\mathrm{sca},\,i}$ efficiencies for the $i$-th grain species are derived from the Mie theory \citep{mie1908} assuming spherical grains.
The refractive indexes of \ce{Al2O3}, \ce{Fe3O4}, \ce{MgSiO3}, \ce{Mg2SiO4}, and \ce{SiO2} are taken from the references shown in Table A of \citet{bianchi2007}. The refractive index of \ce{Fe} is taken from \citet{Palik1991}. We have computed the efficiencies for wavelengths longer than the given range of wavelengths in their table by linear extrapolation.
For \ce{AC} grains, we adopt a model proposed by \citet{jones2013}. In this model, smaller grains $(\lesssim 20\,\mathrm{nm})$ consist of aromatic-rich H-poor amorphous carbon, whereas larger grains have a core/mantle structure (see \citealt{jones2013} for more details\footnote{These extinction properties of carbonaceous grains are available in the \texttt{DustEM} code under the name of \texttt{CM20}.}). 

The extinction efficiency $Q_\mathrm{ext,i}=Q_\mathrm{abs,i}+Q_\mathrm{sca,i}$ depends on grain size $a$ and wavelength $\lambda$. The extinction coefficient normalized to the total dust mass is defined as 
\begin{gather}
    \kappa_\lambda \equiv \frac{1}{M_\mathrm{dust}} \sum_{i\in \mathrm{species}}\int_{a_\mathrm{min}}^{a_\mathrm{max}} \left(\frac{\mathrm{d}n}{\mathrm{d}a}\right)_i \pi a^2 Q_{\mathrm{ext},\,i}(a,\,\lambda)\,\mathrm{d} a,
\end{gather}
and the extinction is calculated as
\begin{gather}
    A_\mathrm{\lambda} = 2.5\, \ln N_\mathrm{H}\sum_{i\in \mathrm{species}}\int_{a_\mathrm{min}}^{a_\mathrm{max}} \left(\frac{\mathrm{d}n}{\mathrm{d}a}\right)_i \pi a^2 Q_{\mathrm{ext},\,i}(a,\,\lambda)\,\mathrm{d} a, \label{eq:ext}
\end{gather}
where $N_\mathrm{H}$ is the column density of hydrogen. 
At long wavelengths, where scattering is negligible, the extinction coefficient approaches the absorption (emissivity) coefficient.
In this study, we normalize the extinction curve using the extinction $A_V$ at the 
V-band wavelength of 5500 \AA{}.
Since the extinction efficiency at 5500 \AA{} for AC grains increases with grain size up to $\sim100\,\mathrm{nm}$ and then nearly reaches a plateau, $A_V$ mainly reflects the abundance of large grains with sizes $\gtrsim100\,\mathrm{nm}$.

\section{Results}
\label{sec:result}
\subsection{Dust mass time evolution}
\label{subsec:massevo}

\begin{figure}[htbp]
    \centering
    \includegraphics[width=\linewidth]{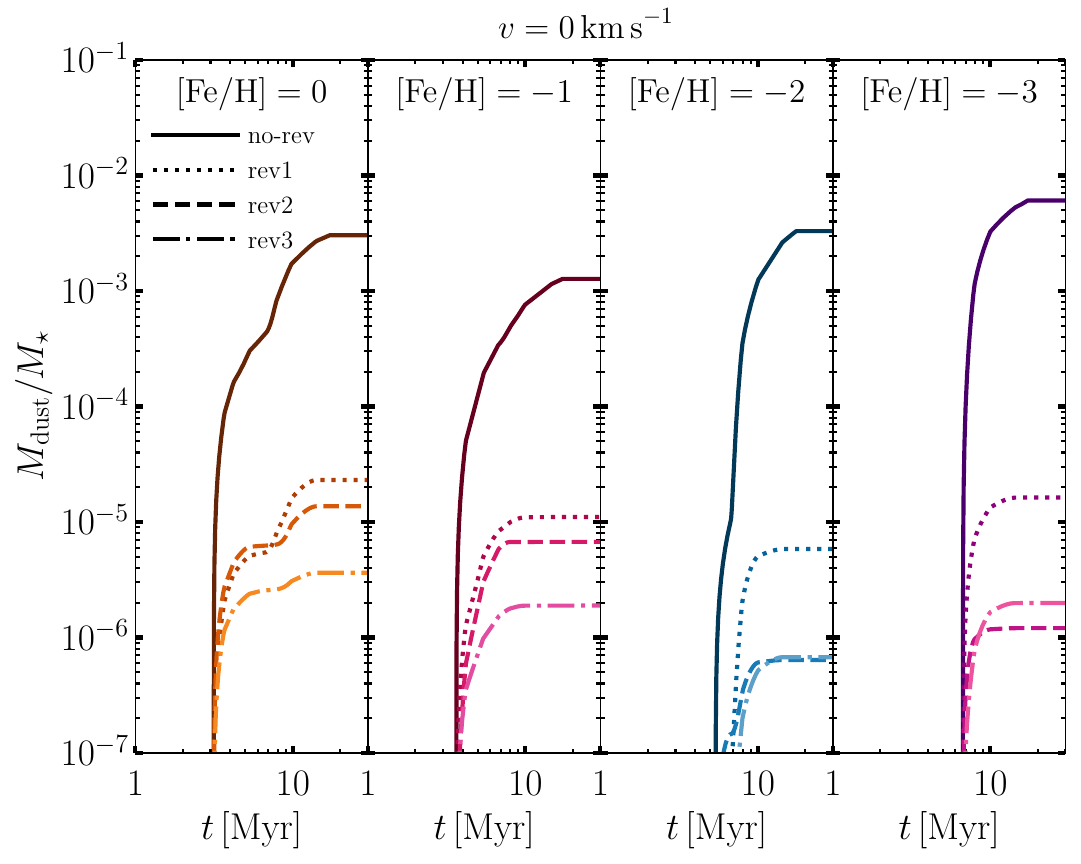}
        \caption{Time evolution of the dust mass released by SNe with nonrotating progenitors, normalized to the total stellar mass formed in a burst at $t=0$. From left to right, the panels show the result for initial stellar metallicities of $[\ce{Fe/H}]=0,\,-1,\,-2$, and $-3$, respectively. The Kroupa IMF is adopted for high metallicity models ($[\ce{Fe/H}]=0,\,-1$, and $-2$), and a log-flat IMF for the lowest metallicity model ($[\ce{Fe/H}]=-3$). In each panel, the solid, dotted, dashed, and dash-dotted lines correspond to the no-reverse shock (no-rev), rev1, rev2, and rev3 models, respectively (see text).
    }
    \label{fig:evol_V0_Kroupa_Logflat}
\end{figure}
\begin{figure}[htbp]
    \centering
    \includegraphics[width=\linewidth]{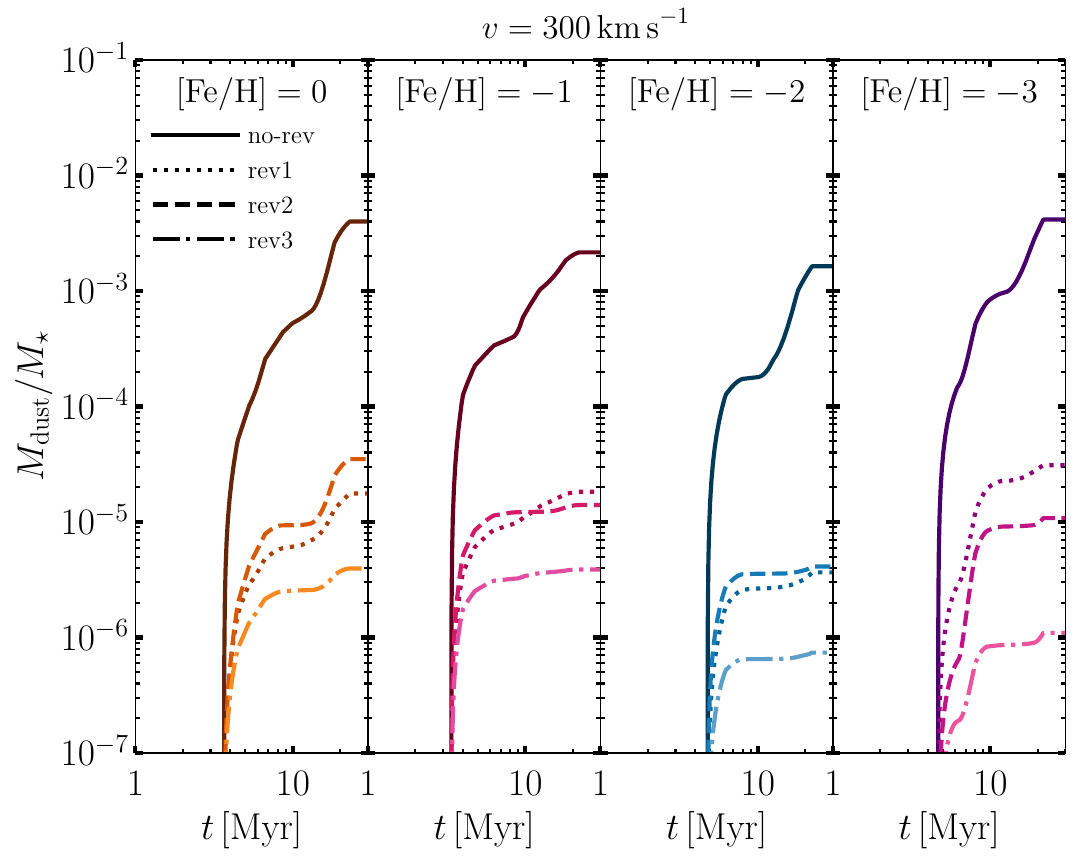}
    \caption{Same as Fig. \ref{fig:evol_V0_Kroupa_Logflat}, but for SNe with rotating progenitors ($v = 300$\, km/s).}
    \label{fig:evol_V300_Kroupa_Logflat}
\end{figure}
\begin{figure}[htbp]
    \centering
    \includegraphics[width=\linewidth]{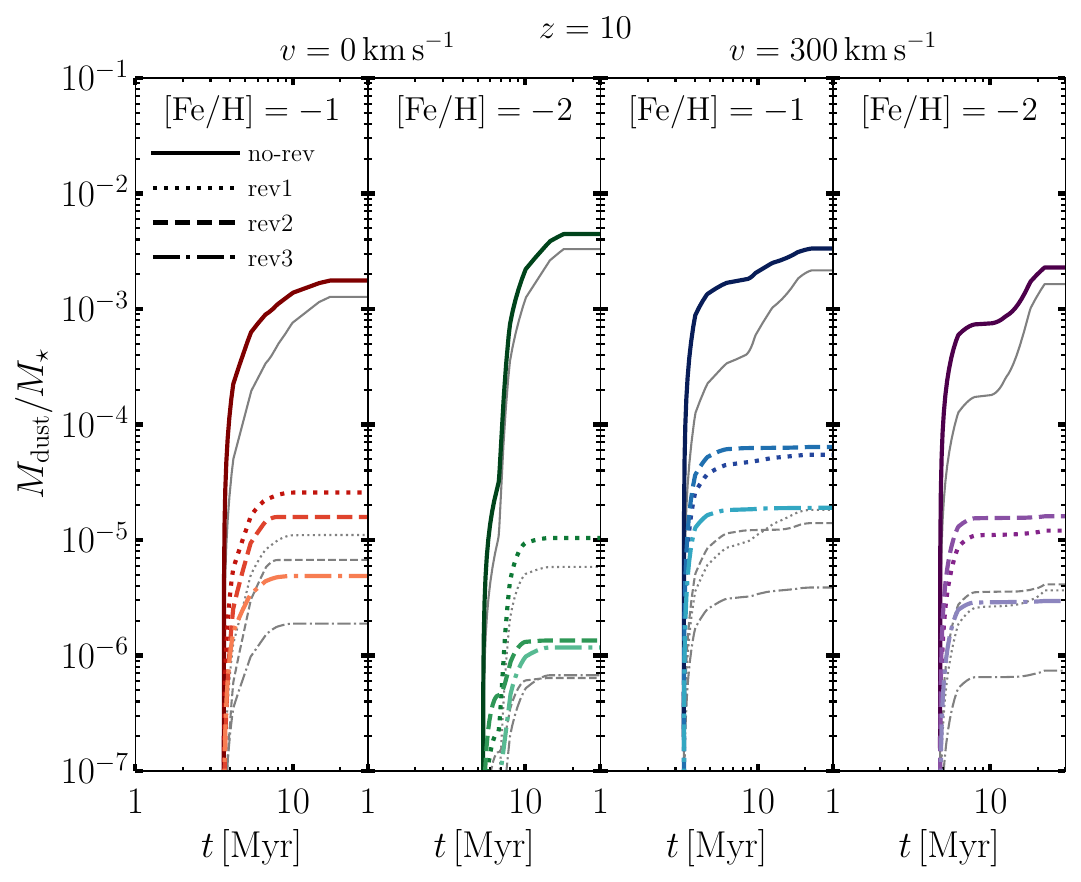}
    \caption{Same as Fig. \ref{fig:evol_V0_Kroupa_Logflat}, but adopting a Chon IMF for redshift $z=10$. From left to right, the panels show results for nonrotating progenitors with $[\ce{Fe/H}]=-1$ and $-2$, and for rotating progenitors with $[\ce{Fe/H}]=-1$ and $-2$. Each thin gray line shows the result obtained assuming a Kroupa IMF, as reported in Figs. \ref{fig:evol_V0_Kroupa_Logflat} and \ref{fig:evol_V300_Kroupa_Logflat}.}
    \label{fig:evol_Chon}
\end{figure}

Figs. \ref{fig:evol_V0_Kroupa_Logflat}-\ref{fig:evol_Chon} show the time evolution, from $t=0$ to $30\,\mathrm{Myr}$, of the total dust mass $M_\mathrm{dust}$ released by SNe, normalized by the total stellar mass $M_\star$ formed at $t=0$.

In Figs. \ref{fig:evol_V0_Kroupa_Logflat} and \ref{fig:evol_V300_Kroupa_Logflat}, the results are shown for SNe with nonrotating and rotating stellar progenitors, respectively. The stars are assumed to form with a Kroupa IMF for $[\ce{Fe/H}]=0,\,-1$, and $-2$, and with a log-flat IMF for $[\ce{Fe/H}]=-3$. 
In each panel, the solid, dotted, dashed, and dash-dotted lines correspond to the no-reverse shock (no-rev) and reverse shock models with $n_\mathrm{ISM}=0.05$ (rev1), $0.5$ (rev2), and $5\,\mathrm{cm^{-3}}$ (rev3). 
For our single stellar population model, in which all the stars are formed in a single burst at $t=0$, the dust mass increases monotonically until the time ($\sim30\,\mathrm{Myr}$ after the starburst) when the least massive star explodes as a supernova.

For the nonrotating models, the destructive effects of the reverse shock reduce the final dust mass by at least 99\% compared to the no-reverse-shock models, and the surviving dust mass generally decreases with increasing ISM density. However, the effective dust yield for each progenitor strongly depends on the initial grain size distribution prior to the passage of the reverse shock.
In some models, the dust yields do not decrease monotonically with increasing ISM density \citep{otaki2026}. Therefore, for $[\ce{Fe/H}]=-2$ and $-3$, the rev3 model yields a total dust mass larger than that of the rev2 model.

Due to the effect of fallback, the maximum progenitor mass that results in a supernova explosion decreases with decreasing metallicity. Consequently, the delay time of dust enrichment following star formation increases at lower metallicity. A saturation in the dust mass evolution is found at $\sim 7\,\mathrm{Myr}$ in the $[\ce{Fe/H}]=0$ and $-2$ cases. This is caused by the inefficient dust production from nonrotating progenitor with $30\, M_\odot$.
At $30\,\mathrm{Myr}$, the maximum and minimum dust masses, accounting for the destructive effects of the reverse shock, are $M_\mathrm{dust}/M_\star\simeq 2.3\times10^{-5}$ in the rev1 model for $[\ce{Fe/H}]=0$, and $M_\mathrm{dust}/M_\star\simeq 6.4\times10^{-7}$ in the rev2 model for $[\ce{Fe/H}]=-2$.

The time evolution of the dust mass released by SNe with rotating progenitors is shown in Figure \ref{fig:evol_V300_Kroupa_Logflat}. The models exhibit a trend similar to that of 
nonrotating models. However, the rev2 models with $[\ce{Fe/H}]=0$ and $-2$ consistently yield larger dust masses than the rev1 models, because the massive progenitors ($\gtrsim 40\, M_\odot$) in the rev2 model produce more dust mass than those in the rev1 model. 
We note that dust grains are not uniformly destroyed during the passage of the reverse shock, as discussed in \citet{bocchio2016}. Since the efficiency of destruction caused by the reverse shock strongly depends on the initial grain size distribution prior to its passage, the largest effective dust yields are not always associated to SNe exploding in the lowest-density ISM \citep{otaki2026}. Compared to the nonrotating models, a saturation in the dust mass evolution at $\sim 7\,\mathrm{Myr}$ is apparent due to the lower contribution from $20$--$25\,M_\odot$ progenitors in the no-rev model.
In reverse shock models, the maximum and minimum dust masses at $30\,\mathrm{Myr}$ are $M_\mathrm{dust}/M_\star\simeq 3.5\times10^{-5}$ in the rev2 model for $[\ce{Fe/H}]=0$, and $M_\mathrm{dust}/M_\star\simeq 7.4\times10^{-7}$ in the rev3 model for $[\ce{Fe/H}]=-2$.

Figure \ref{fig:evol_Chon} shows how the dust mass evolution changes when we adopt the Chon IMF. Here we assume a specific redshift of $z=10$. The left and right pairs of panels represent nonrotating and rotating models, respectively. In each model, the left and right sides indicate $[\ce{Fe/H}]=-1$ and $-2$.
The gray lines indicate the corresponding evolution expected for a Kroupa IMF.
Due to the higher fraction of massive stars in the Chon IMF relative to the Kroupa IMF, the total dust masses are systematically higher and increase more rapidly at early times ($\lesssim 10\,\mathrm{Myr}$). In particular, since more massive progenitors generally produce larger grains, which are more resistant to the destructive effects of the reverse shock, the difference between the final dust masses obtained with the Chon and Kroupa IMFs tends to be larger in the rev models than in the no-rev models.
For the nonrotating and rotating models that include the destructive effects of the reverse shock, the maximum dust masses at $30\,\mathrm{Myr}$ are $M_\mathrm{dust}/M_\star \simeq 2.6 \times 10^{-5}$ and $6.4 \times 10^{-5}$ in the rev1 and rev2 models with $[\ce{Fe/H}] = -1$, respectively. The minimum dust masses correspond to $M_\mathrm{dust}/M_\star \simeq 1.2 \times 10^{-6}$ and $3.0 \times 10^{-6}$ in the rev3 model with $[\ce{Fe/H}] = -2$ for the nonrotating and rotating progenitors, respectively.

\subsection{Extinction curves for models with no reverse shock}
\begin{figure}
    \centering
    \includegraphics[width=\linewidth]{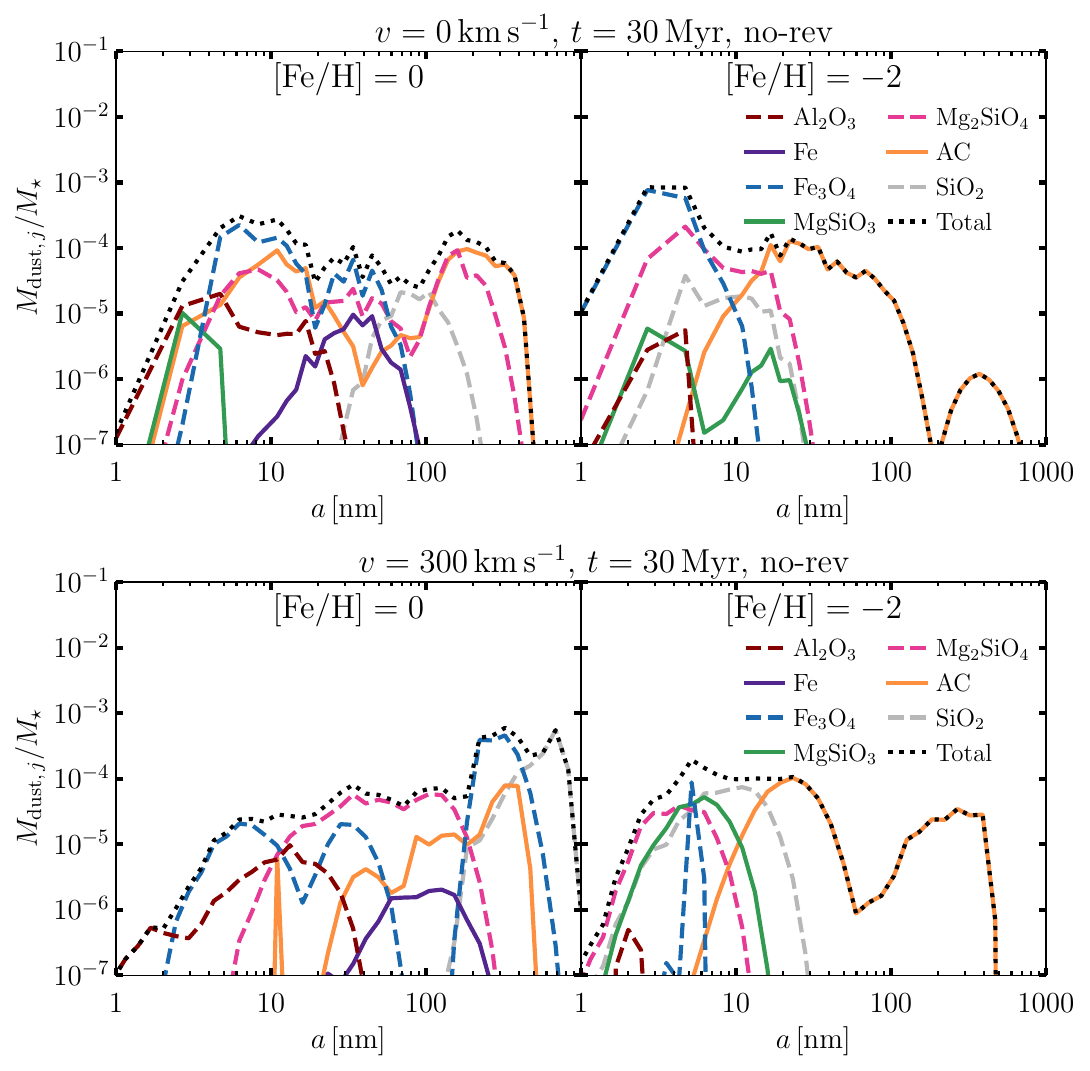}
    \caption{Grain size distributions for different dust species in the no-reverse-shock models at $30\,\mathrm{Myr}$ assuming a Kroupa IMF. Each panel corresponds to different progenitor properties: nonrotating stars with $[\ce{Fe/H}] = 0$ (top left), nonrotating stars with $[\ce{Fe/H}] = -2$ (top right), rotating stars with $[\ce{Fe/H}] = 0$ (bottom left), and rotating stars with $[\ce{Fe/H}] = -2$ (bottom right). The solid and dashed lines show the contribution of \ce{Al2O3} (red), \ce{Fe} (purple), \ce{Fe3O4} (blue), \ce{MgSiO3} (green), \ce{Mg2SiO4} (pink), \ce{AC} (orange), and \ce{SiO2} (gray). The black dotted line indicates the total size distribution.}
    \label{fig:distri_comp}
\end{figure}
\begin{figure}
    \centering
    \includegraphics[width=0.95\linewidth]{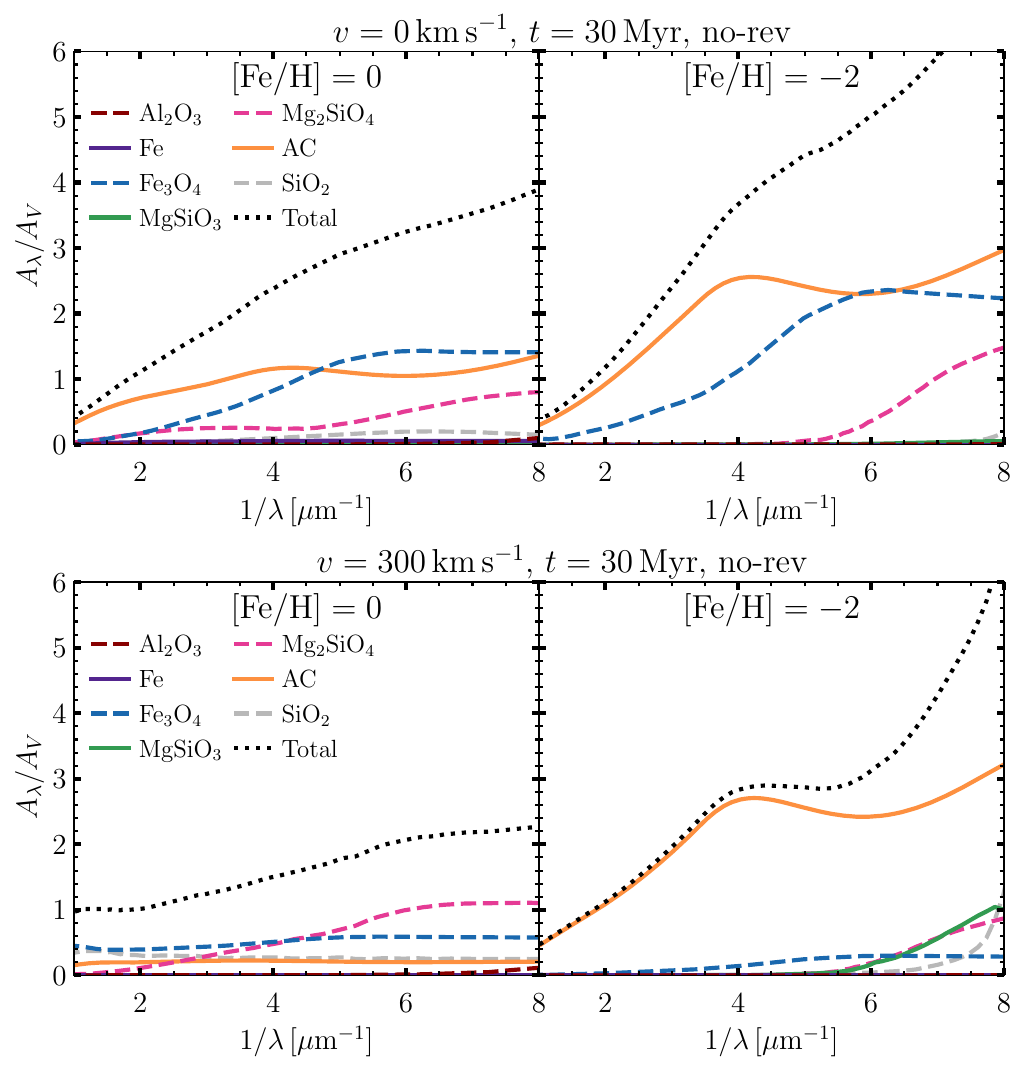}
    \caption{Contribution of each grain species to the total extinction curve adopting the same models shown in Fig. \ref{fig:distri_comp}. The lines follow the same colour-coding and line-styles adopted in Fig. \ref{fig:distri_comp}.}
    \label{fig:ext_comp}
\end{figure}

Figure~\ref{fig:distri_comp} shows the grain size distributions for each grain species at $30\,\mathrm{Myr}$ in the no-rev models with the Kroupa IMF. The vertical axis indicates the dust mass in each size bin, $M_{\mathrm{dust},j}$, normalized by the total stellar mass $M_\star$. Each panel corresponds to a different model: nonrotating stars with $[\ce{Fe/H}] = 0$ (top left) and $[\ce{Fe/H}] = -2$ (top right), and rotating stars with $[\ce{Fe/H}] = 0$ (bottom left) and $[\ce{Fe/H}] = -2$ (bottom right). In each panel, the solid and dashed lines correspond to the distributions of \ce{Al2O3} (red), \ce{Fe} (purple), \ce{Fe3O4} (blue), \ce{MgSiO3} (green), \ce{Mg2SiO4} (pink), \ce{AC} (orange), and \ce{SiO2} (gray). 
The black dotted line indicates the total grain size distribution. In both the nonrotating and rotating models, various species are distributed over a wide range of grain sizes for $[\ce{Fe/H}] = 0$, whereas for $[\ce{Fe/H}] = -2$, AC grains dominate the size distribution at larger sizes, with $a \gtrsim 10\,\mathrm{nm}$.

For nonrotating stars with $[\ce{Fe/H}] = 0$, \ce{Fe3O4} grains have the highest mass fraction, $0.36$, followed by \ce{AC} and \ce{Mg2SiO4} with mass fractions of $0.34$ and $0.21$, respectively. Similarly, in the $[\ce{Fe/H}] = -2$ model, \ce{Fe3O4}, \ce{AC}, and \ce{Mg2SiO4} grains are also the dominant contributors to the dust mass, with mass fractions of $0.45$, $0.32$, and $0.19$, respectively.

On the other hand, the rotating model with $[\ce{Fe/H}] = 0$ shows mass fractions of $0.43$, $0.33$, and $0.14$ for \ce{Fe3O4}, \ce{SiO2}, and \ce{Mg2SiO4}, respectively. For the $[\ce{Fe/H}] = -2$ rotating model, the mass contributions of \ce{AC}, \ce{SiO2}, and \ce{MgSiO3} are significant, with mass fractions of $0.39$, $0.28$, and $0.15$, respectively.

Figure~\ref{fig:ext_comp} shows the extinction curves at $30\,\mathrm{Myr}$ calculated using Eq. \eqref{eq:ext} for the same set of models shown in Figure \ref{fig:distri_comp}. The vertical axis indicates the extinction $A_\lambda$ normalized by the extinction $A_V$ in the $V$-band, and the horizontal axis represents the inverse of the wavelength $1/\lambda$.

In general, the $[\ce{Fe/H}] = -2$ models exhibit a steeper total extinction curve than the $[\ce{Fe/H}] = 0$ models due to the contribution of relatively small AC grains ($\simeq10\text{--}100\,\mathrm{nm}$), broadly reflecting the mass fractions of grain species shown in Figure \ref{fig:distri_comp}. In particular, AC grains produce a broad bump at $1/\lambda \simeq 4\,\mu\mathrm{m}^{-1}$, corresponding to $\lambda \simeq 2500\,\text{\AA}$. This is reflected into a distinct bump in the total extinction curve for the rotating model with $[\ce{Fe/H}] = -2$. However, in the nonrotating model with the same initial metallicity, the large contributions from \ce{Fe3O4} and \ce{Mg2SiO4} result in a steep extinction curve without a bump.

\subsection{Extinction curves for models with reverse shock}

\begin{figure*}[htbp]
    \centering
    \includegraphics[width=0.8\linewidth]{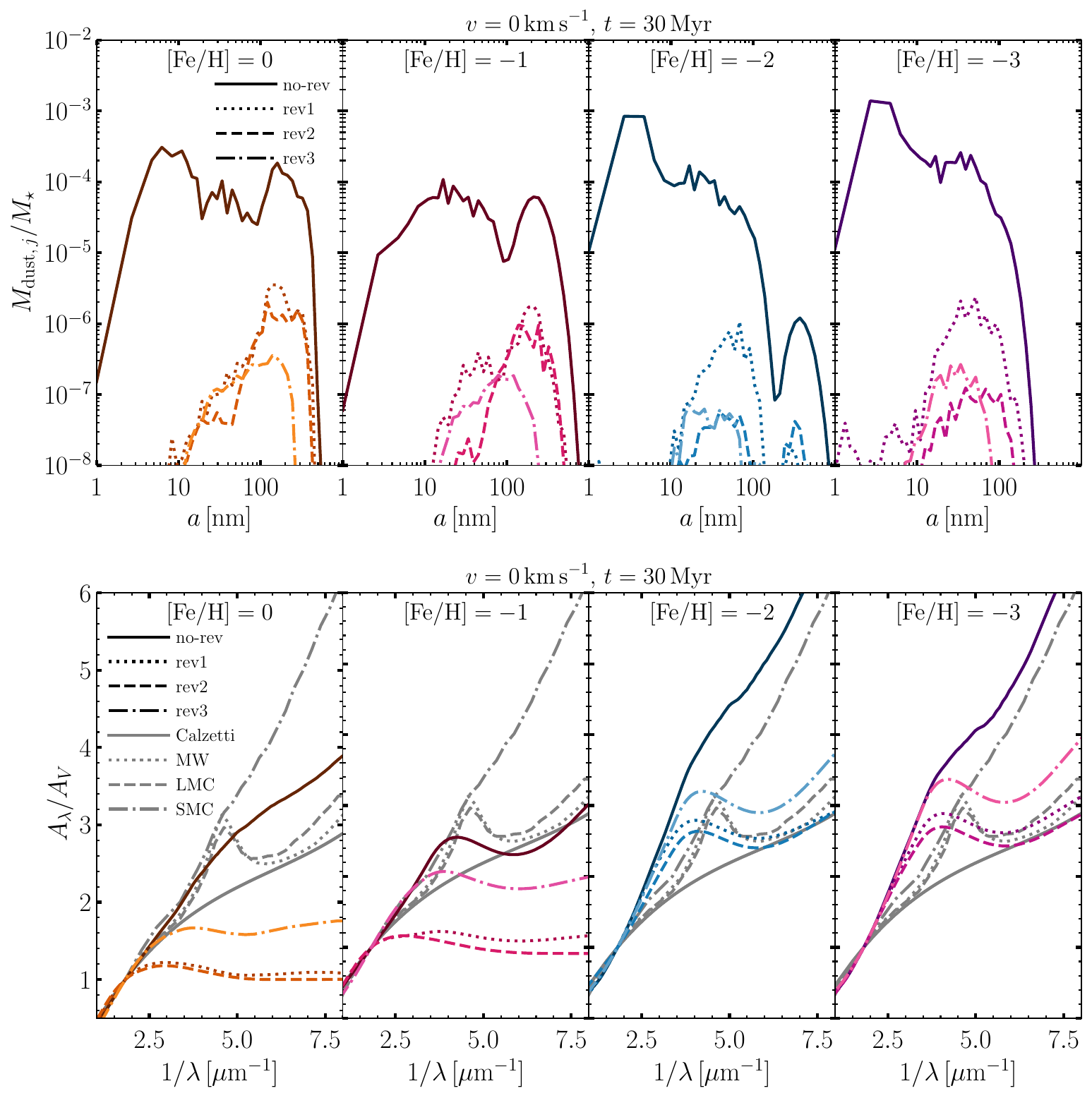}
    \caption{Grain size distributions (top panels) and extinction curves (bottom panels) for nonrotating models at $30\,\mathrm{Myr}$. From left to right, the panels show results for initial stellar metallicity of $[\ce{Fe/H}] = 0$, $-1$, $-2$, and $-3$, respectively. The Kroupa IMF is adopted for the high-metallicity models ($[\ce{Fe/H}] = 0$, $-1$, and $-2$), while a log-flat IMF is used for the lowest-metallicity one ($[\ce{Fe/H}] = -3$). In the top panels, the horizontal axis represents the grain size, while the vertical axis indicates the corresponding dust mass normalized by the stellar mass. The solid, dotted, dashed, and dash-dotted lines correspond to the no-reverse-shock (no-rev), rev1, rev2, and rev3 models, respectively. In the bottom panels, the horizontal axis is the inverse wavelength, and the vertical axis shows the dust extinction normalized by the $V$-band extinction. The gray solid, dotted, dashed, and dash-dotted lines correspond to the attenuation curves of \citet{calzetti2000}, the Milky Way \citep{pei1992}, the LMC average \citep{gordon2003}, and the SMC bar \citep{gordon2003}, respectively.}
    \label{fig:distri_ext_V0_Kroupa_Logflat}
\end{figure*}

\begin{figure*}[htbp]
    \centering
    \includegraphics[width=0.8\linewidth]{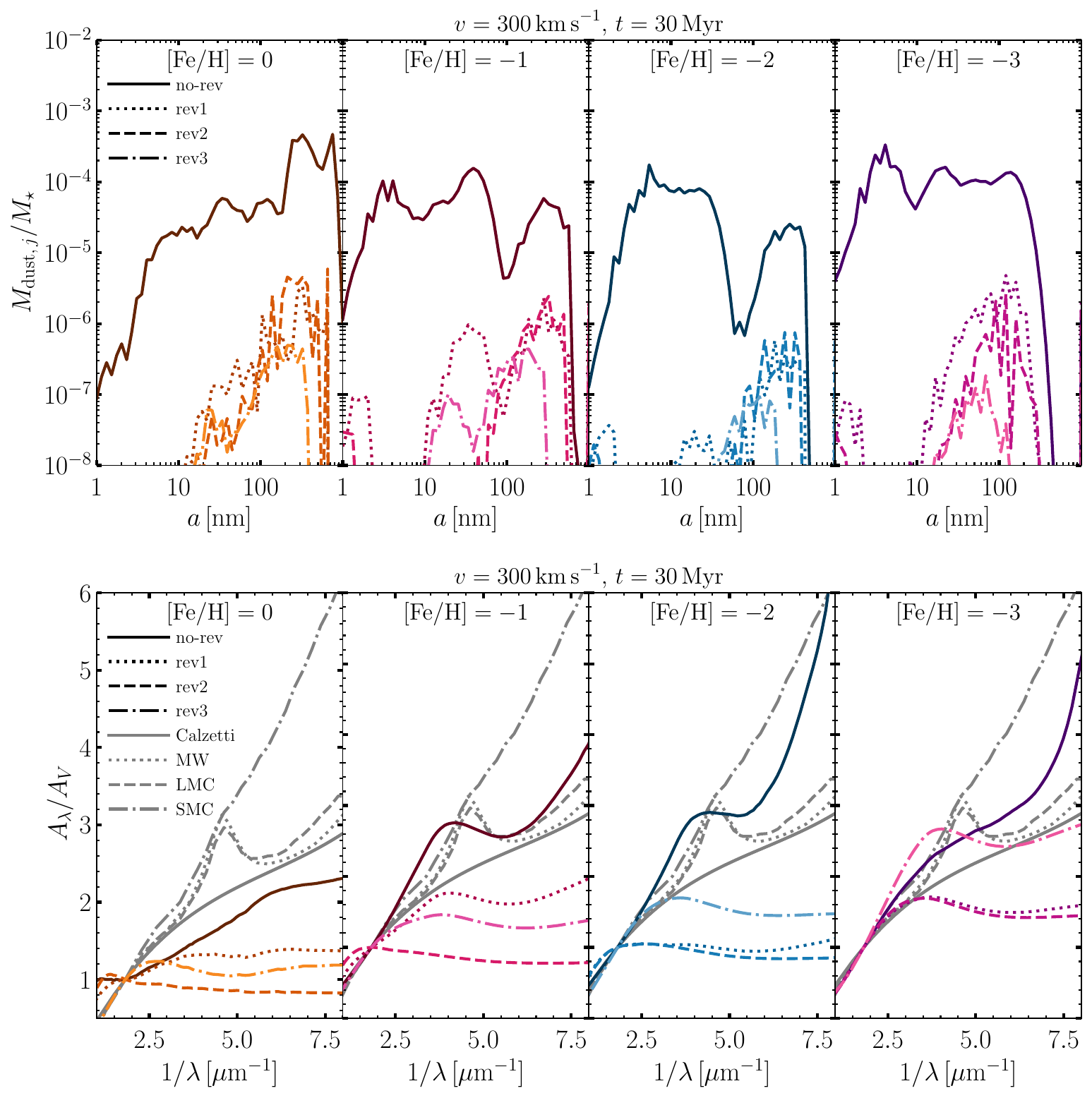}
    \caption{Same as Fig. \ref{fig:distri_ext_V0_Kroupa_Logflat}, but for rotating models.}
    \label{fig:distri_ext_V300_Kroupa_Logflat}
\end{figure*}

In Figure~\ref{fig:distri_ext_V0_Kroupa_Logflat}, the top and bottom panels show the grain size distributions and extinction curves, respectively, for nonrotating models at $t = 30\,\mathrm{Myr}$. From left to right, each panel corresponds to a different initial stellar metallicity: $[\ce{Fe/H}] = 0$, $-1$, $-2$, and $-3$. The Kroupa IMF is adopted for the high-metallicity models ($[\ce{Fe/H}] = 0$, $-1$, and $-2$), while a log-flat IMF is used for the lowest-metallicity model ($[\ce{Fe/H}] = -3$). The solid, dotted, dashed, and dash-dotted lines correspond to the no-reverse-shock (no-rev), rev1, rev2, and rev3 models, respectively. 

In the bottom panels, the gray solid, dotted, dashed, and dash-dotted lines represent the observed attenuation curves of local starbursts \citep{calzetti2000}, the MW \citep{pei1992}, the LMC, and the SMC \citep{gordon2003}, respectively.

For all nonrotating models, the reverse shock reduces the mass of small grains with sizes $\lesssim 10\,\mathrm{nm}$. Most of the surviving large grains are composed of AC. The reverse-shock models systematically produce flatter extinction curves than the no-reverse-shock models, reflecting the modified grain size distributions. For $[\ce{Fe/H}] = 0$ and $-1$, the extinction curves become flatter than most of the observed ones.

At $[\ce{Fe/H}] = 0$, the flatter slope is mainly driven by the contribution of species other than AC, whereas the lower-metallicity models show steeper curves with a broad bump due to the increased abundance of AC grains relative to the other species. Therefore, as metallicity decreases, the slope of the extinction curve becomes steeper than the Calzetti curve after the passage of the reverse shock. In the lower-metallicity cases ($[\ce{Fe/H}] = -2$ and $-3$), the extinction curves of the reverse-shock models are similar to those of the MW, LMC, and Calzetti curves.

In general, small carbonaceous grains with sizes $\lesssim 1\,\mathrm{nm}$ are expected to produce the $2175\,\text{\AA}$ bump. However, in our nonrotating models, these small grains are efficiently destroyed by the reverse shock. As a result, a broad bump dominated by larger AC grains appears at $\sim 2500\,\text{\AA}$, even in the no-reverse-shock models. This feature was previously reported by \citet{bianchi2007}.

Figure~\ref{fig:distri_ext_V300_Kroupa_Logflat} shows the results for the rotating models. As in the nonrotating case, the reverse shock efficiently destroys small grains with sizes $\lesssim 10\,\mathrm{nm}$. However, for $[\ce{Fe/H}] = -2$ and $-3$, the extinction curves are flatter than those of the corresponding nonrotating models.

A notably steep slope appears in the $[\ce{Fe/H}] = -3$ rev3 model, which exhibits a relatively high AC fraction compared to the other reverse-shock models. This behavior indicates that the other grain species are more efficiently destroyed by the reverse shock, thereby enhancing the relative contribution of AC and steepening the curve.

Overall, the rotating models tend to produce flatter extinction curves than both the nonrotating models and the Calzetti law. This difference arises because rotating progenitors form larger grains ($\gtrsim 10\,\mathrm{nm}$) in the ejecta before the reverse shock propagates through them. These larger grains are more resistant to destruction and therefore preserve a flatter extinction curve.

We also examine the time-dependent evolution of the predicted extinction curves, consistently with the dust mass evolution discussed in Section 4.1. To illustrate how the grain size distributions and the resulting extinction curves change over time, we present in Appendix A the results at three representative epochs after the burst: $t = 7$, 10, and $30 \,\mathrm{Myr}$ (see Figures A.1–A.4).
The timescale of $7 \,\mathrm{Myr}$ corresponds approximately to the lifetimes of nonrotating and rotating progenitors with stellar masses of $m_\star \simeq 25\,M_\odot$ and $30\,M_\odot$, respectively. Therefore, at this early stage, dust production is dominated by the most massive progenitors, with initial masses $\gtrsim 25$–$30\,M_\odot$. 

Overall, the time-dependent analysis confirms that the evolution of extinction curves is primarily driven by the progressive contribution of less massive supernova progenitors and by the efficiency of reverse-shock destruction. While no-reverse-shock models generally develop steeper slopes over time due to the increasing abundance of small grains, reverse-shock models show much weaker temporal evolution and tend to preserve flatter extinction curves, especially in the rotating progenitor cases.

\section{Comparison with observations }
\label{sec:discuss}

\subsection{Extinction properties}

\begin{figure*}
    \centering
    \includegraphics[width=0.49\linewidth]{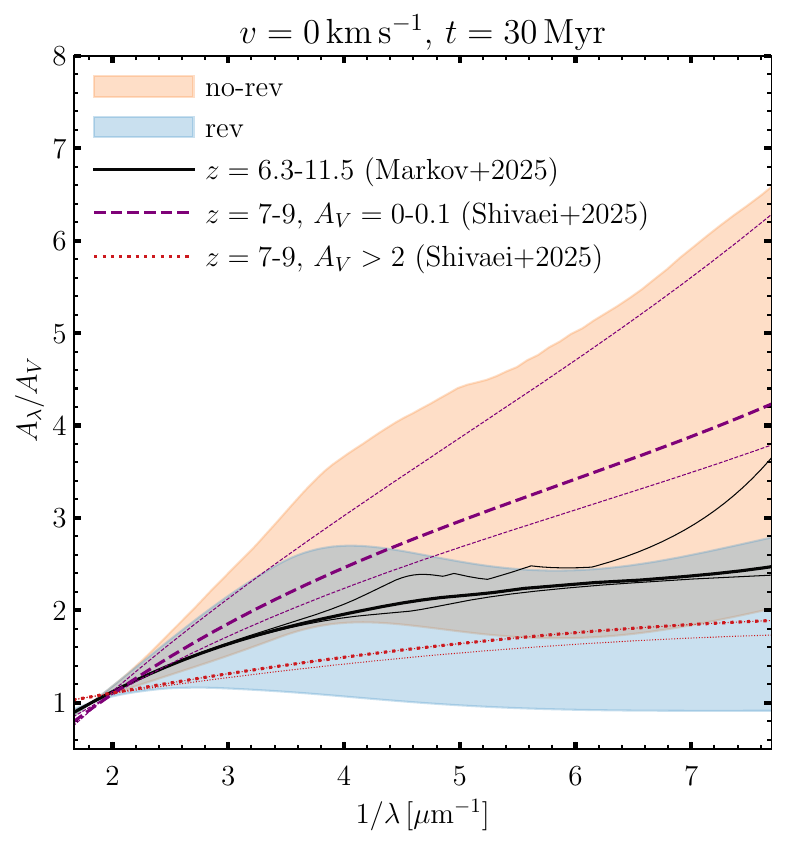}
    \includegraphics[width=0.49\linewidth]{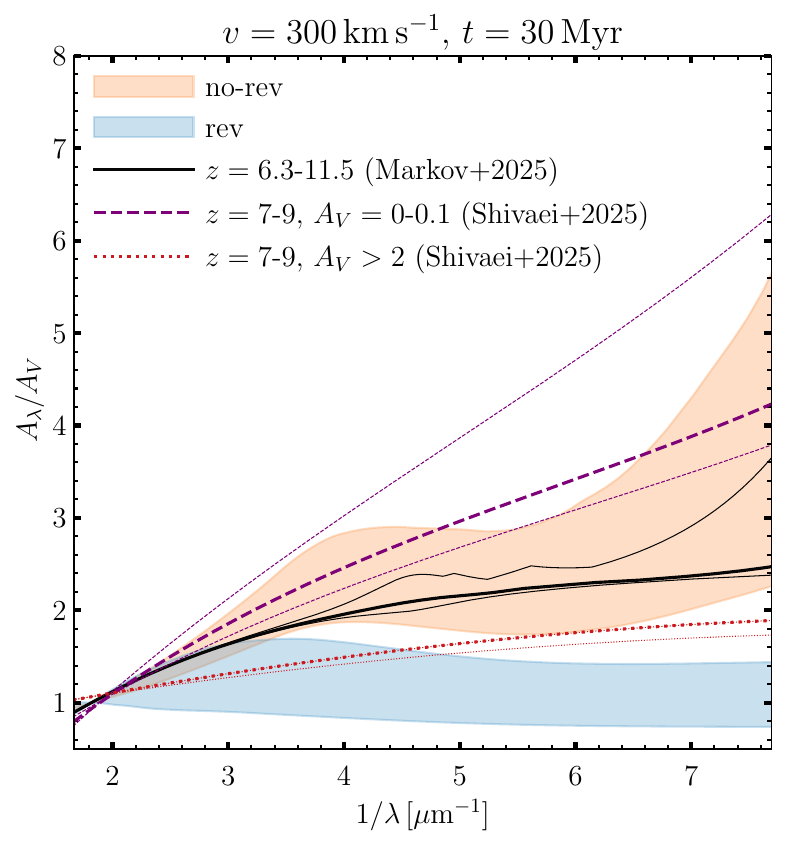}
    \caption{Comparison between our theoretical extinction curves and observed attenuation curves. The left and right panels indicate the results of the nonrotating and rotating models. The orange and blue shaded regions correspond to the range of extinction curves predicted in the no-rev and rev2 models (indicated as rev in the legend of the figures), spanning variations in metallicity ($[\ce{Fe/H}]=0,\, -1,\, -2,$ and $-3$) and IMF (Kroupa and Chon).
    The thick solid line represents the median attenuation curve derived by \citet{Markov2025} at $z=6.3$-$11.5$, with thinner black lines indicating the $1-\sigma$ error.  The purple dashed and red dotted lines correspond to the average observed attenuation curves (with thinner lines indicating the 16th–84th percentile range) derived by \citet{shivaei2025} at $z=7$-$9$ for two subsamples of galaxies with $A_V=0$-$0.1$ and $A_V>2$, respectively.}
    \label{fig:ext_obs}
\end{figure*}

The results presented in the previous section describe the time evolution of dust properties originating from supernova explosions, which are expected to trace the dust properties of high-redshift galaxies owing to the short timescales of dust production \citep{valiante2009}. In this section, we compare our results with recent JWST-based studies of the attenuation properties of high-redshift galaxies.

Figure~\ref{fig:ext_obs} shows a comparison between the observed attenuation curves and our model predictions for the nonrotating (left) and rotating (right) cases. In each panel, the orange and blue shaded regions represent the range of extinction curves predicted by the no-reverse-shock (no-rev) and rev2 (indicated as rev in the legend) models, respectively, spanning variations in metallicity ($[\ce{Fe/H}]=0,-1,-2$, and $-3$) and IMF (Kroupa and Chon). The rev2 model is adopted because it corresponds to the case in which dust destruction by the reverse shock is most efficient. Therefore, the no-rev and rev2 models can be regarded as two limiting scenarios, bracketing the range of possible extinction curves.
The solid line corresponds to the median attenuation curve provided by \citet{Markov2025} for galaxies at $z = 6.3$–11.5. The dashed and dotted lines show the average attenuation curves reported by \citet{shivaei2025} for galaxies at $z = 7$–9, divided by different levels of V-band attenuation, namely $A_V = 0$–0.1 and $A_V > 2$, respectively.

When comparing our model predictions with these observational constraints, it is important to keep in mind the distinction between attenuation and extinction curves. The attenuation curves of high-redshift galaxies are derived from observed spectral energy distributions by adopting analytical parameterizations, such as the Calzetti law. In contrast, extinction curves describe the intrinsic absorption and scattering properties of dust. Therefore, attenuation curves additionally incorporate geometrical and radiative transfer effects.

We find that most of our theoretical extinction curves are broadly consistent with the observed attenuation curves at high redshift. In particular, the flat attenuation curves with $A_V > 2$ observed at $z = 7$--$9$ are well reproduced by the nonrotating models that include the destructive effects of the reverse shock.
This consistency may indicate that, for at least part of the high-redshift galaxy population, the effective dust-star geometry is relatively simple. However, this interpretation is not unique, as different combinations of dust geometry, optical depth, and scattering can produce similar attenuation curves.

Conversely, the range of extinction curves predicted by the no-reverse-shock (no-rev) models is consistent with the low-attenuation case, $A_V = 0$--0.1. These observational trends may reflect galaxies in which the reverse shock is less effective, for instance, in low-density regions of the ISM, or systems where additional dust processing mechanisms, such as shattering or grain growth, play an important role. Such effects are not included in the present work.

For rotating progenitor models, grains with initial sizes $\gtrsim 10\,\mathrm{nm}$ formed during nucleation prior to the passage of the reverse shock are more resistant to destruction. As a result, the larger surviving grain population increases the overall attenuation and produces flatter extinction curves.

Moreover, our models show evidence for a broad carbonaceous bump, in qualitative agreement with recent observational indications reported by \citet{witstok2023b} and \citet{markov2023, Markov2025}, suggesting that carbon-rich dust may already contribute to the attenuation properties of galaxies at these early epochs.

Finally, since our calculations do not account for the complex geometry of galaxies, a more detailed and quantitative comparison with observations will require dedicated radiative transfer simulations in future work.

\subsection{Emissivity properties}

\begin{figure*}
    \centering
    \includegraphics[width=0.7\linewidth]{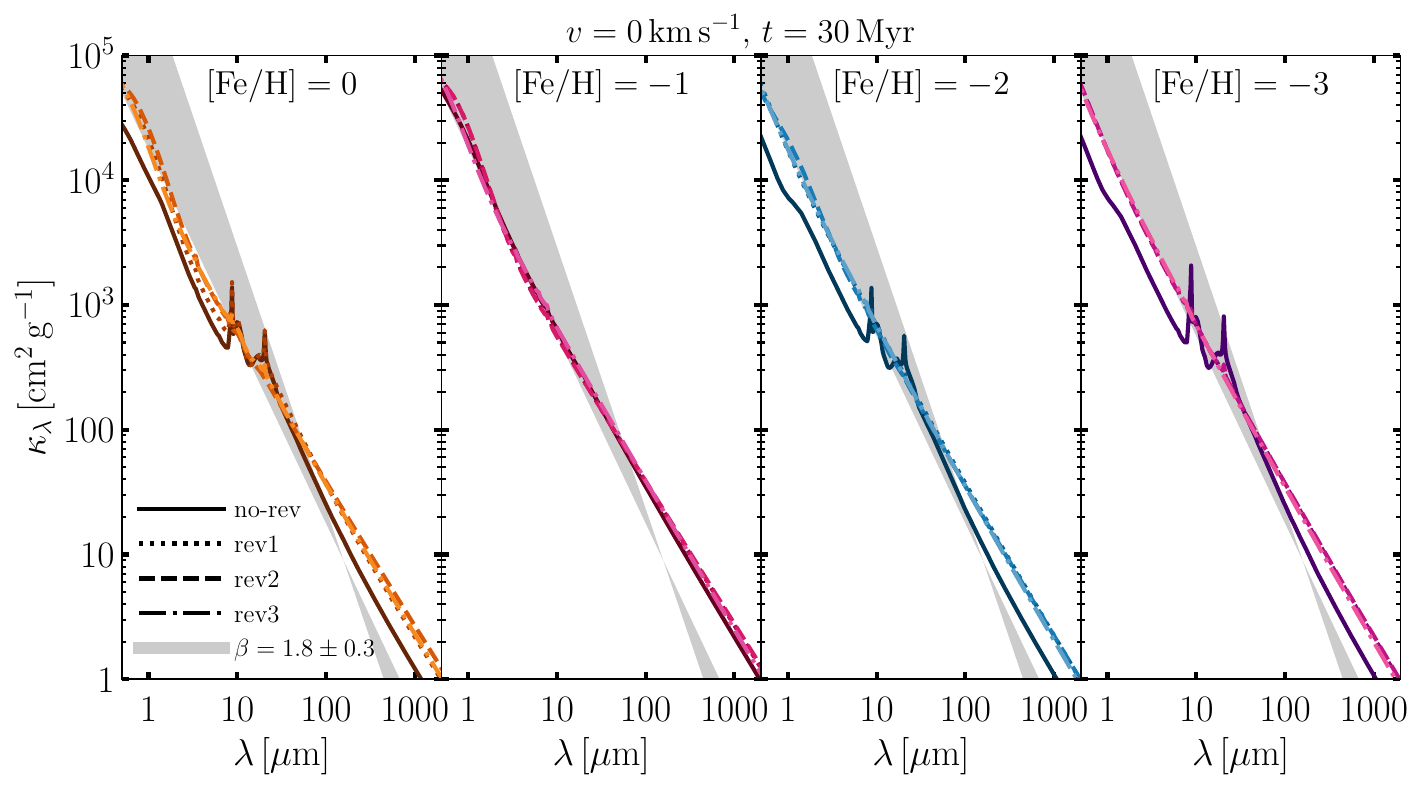}
    \caption{Emissivity per unit dust mass as a function of wavelength for nonrotating models at $t = 30\,\mathrm{Myr}$. From left to right, the panels show results for initial stellar metallicity of $[\ce{Fe/H}] = 0, -1, -2$, and $-3$, respectively. The Kroupa IMF is adopted for the high-metallicity models ($[\ce{Fe/H}] = 0, -1$, and $-2$), while a log-flat IMF is used for the lowest-metallicity model ($[\ce{Fe/H}] = -3$).
    The solid, dotted, dashed, and dash-dotted lines correspond to the no-reverse-shock (no-rev), rev1, rev2, and rev3 models, respectively. The gray shaded region indicates the range of the mean dust emissivity index inferred for high-redshift galaxies at $4 < z < 8$ as reported by \citet{witstok2023}.}
    \label{fig:emiss_V0}
\end{figure*}

\begin{figure*}
    \centering
    \includegraphics[width=0.7\linewidth]{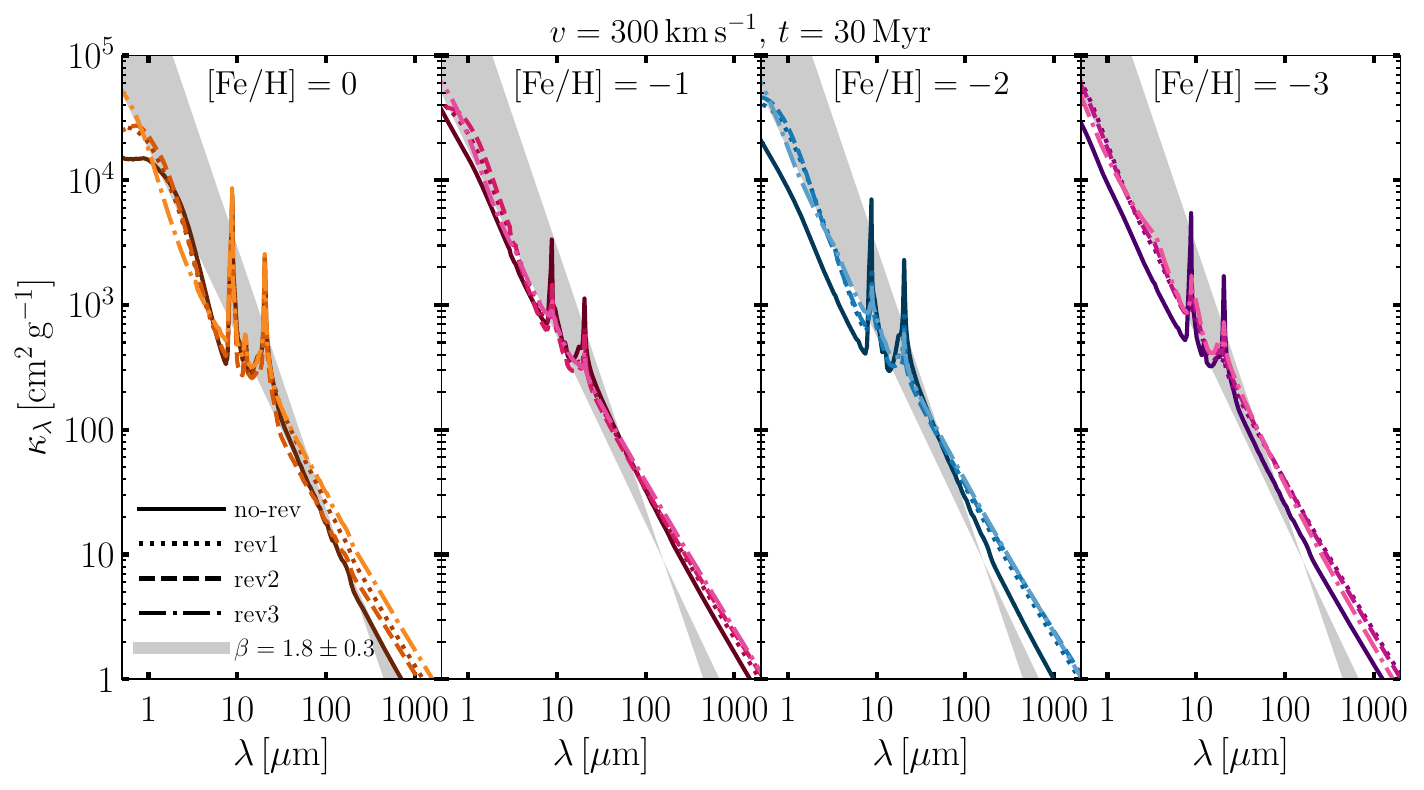}
    \caption{Same as Fig. \ref{fig:emiss_V0}, but for rotating models.}
    \label{fig:emiss_V300}
\end{figure*}

Using the same approach adopted in the calculation of the extinction curves, we use the predicted time-dependent dust masses and grain size distributions from our model grid to compute the emissivity properties expected for galaxies dominated by young stellar populations, in which dust enrichment is primarily driven by supernovae.

We show the emissivity normalized by the total dust mass as a function of wavelength for the nonrotating and rotating models in Figs. \ref{fig:emiss_V0} and \ref{fig:emiss_V300}, respectively. 
The Kroupa IMF is adopted for the high-metallicity models ([Fe/H] = 0, -1, and -2), while a log-flat IMF is used for the lowest-metallicity model ([Fe/H] = -3). The solid, dotted, dashed, and dash-dotted lines correspond to the no-reverse-shock (no-rev), rev1, rev2, and rev3 models, respectively. 
In each panel, the gray shaded region corresponds to the range of the mean dust emissivity index, $\beta_{\rm IR} = 1.8 \pm 0.3$, inferred for observed galaxies at $4 < z < 8$ \citep[][see also the compilation and recent measurements by \citealt{algera2024b}]{witstok2023}. The observed emissivity is parametrized \citep[e.g.,][]{Casey2012, Casey2014} as:
\begin{equation}
\kappa_\nu = \kappa_{\nu,\mathrm{ref}} \left( \frac{\nu}{\nu_{\mathrm{ref}}} \right)^{\beta_{\rm IR}}, \label{eq:kappa}
\end{equation}
where $\kappa_{\nu,\mathrm{ref}} = 8.94\,\mathrm{cm^2\,g^{-1}}$ at $\lambda_{\mathrm{ref}} = 158\,\mu\mathrm{m}$ \citep{hirashita2014,schouws2022,witstok2022}.

For the nonrotating progenitors, the no-reverse-shock (no-rev) models—except for the $[\ce{Fe/H}] = -1$ case—exhibit prominent silicate features at $\lambda \simeq 10\,\mu$m and $20\,\mu$m. The strength of these features depends on the mass fraction of silicate grains formed in the ejecta and on their survival after the passage of the reverse shock.

Among the reverse-shock models, such silicate features are only present in the rev1 model with $[\ce{Fe/H}] = 0$, while they disappear in the other cases due to more efficient dust destruction. The passage of the reverse shock steepens the emissivity in the near-IR (1–10 $\mu$m), but makes it shallower in the far-IR ($\gtrsim 100\,\mu$m) compared to the no-rev models.

In general, the steeper near-IR slope predicted by the reverse-shock models is consistent with observations. However, the far-IR emissivities of our SN-dust models tend to be higher than those inferred observationally, implying smaller dust masses for a given observed flux density in high-redshift galaxies. It is important to note that the observationally inferred emissivity is derived by fitting a modified black body to the observed emission. Temperature mixing due to different sizes and materials can make this apparent $\beta$ different from the effective $\beta$ of a dust model (see e.g. \citealt{hunt2015, bianchi2018, bianchi2022}). Hence, the above comparison should be taken with caution.

For the rotating progenitor models, similar conclusions hold regarding the overall modification of the emissivity slope. However, in this case the silicate features persist even after the passage of the reverse shock, largely independent of metallicity. This suggests that silicate grains produced in rotating progenitors are more resilient and can survive the destructive effects of the reverse shock more efficiently than in the nonrotating case.

Finally, we derive the emissivity parameters $\kappa_{\nu,\mathrm{ref}}$ and $\beta_{\mathrm{IR}}$ by fitting Eq.~(13) over the full wavelength range. The best-fit values for each model are listed in Table 1. In most cases, the emissivity coefficient $\kappa_{\nu,\mathrm{ref}}$ exceeds the reference value of $8.94\,\mathrm{cm}^2\,\mathrm{g}^{-1}$, while the emissivity index $\beta_{\mathrm{IR}}$ is smaller than the observed value of $1.8 \pm 0.3$. This behavior reflects the dominance of larger grains in our models.

\begin{table}[htbp]
\centering
\caption{Fitting parameters of emissivity for all models. The different columns report the rotation velocity of the progenitor star $v$, the metallicity $[\ce{Fe/H}]$, the interstellar medium density $n_\mathrm{ISM}$ (the dash denotes a no-reverse shock model), the coefficient of emissivity $\kappa_{\nu,\,\mathrm{ref}}$, and the emissivity index $\beta_\mathrm{IR}$.}
\label{tab:emissivity}
\begin{tabular}{cllll}
$v\,\mathrm{[km\,s^{-1}]}$	&$[\ce{Fe/H}]$	&$n_\mathrm{ISM}\,\mathrm{[cm^{-3}]}$	&$\kappa_
{\nu,\,\mathrm{ref}}\,\mathrm{[cm^2\,g^{-1}]}$	&$\beta_\mathrm{IR}$	\\
\hline\hline
$0$	&$0$	&$-$	&$11.97$	&$1.22$	\\
$0$	&$0$	&$0.05$	&$16.35$	&$1.04$	\\
$0$	&$0$	&$0.5$	&$18.27$	&$1.03$	\\
$0$	&$0$	&$5$	&$13.43$	&$1.14$	\\
$0$	&$-1$	&$-$	&$12.64$	&$1.20$	\\
$0$	&$-1$	&$0.05$	&$13.44$	&$1.09$	\\
$0$	&$-1$	&$0.5$	&$16.12$	&$1.04$	\\
$0$	&$-1$	&$5$	&$15.09$	&$1.13$	\\
$0$	&$-2$	&$-$	&$11.57$	&$1.28$	\\
$0$	&$-2$	&$0.05$	&$12.66$	&$1.18$	\\
$0$	&$-2$	&$0.5$	&$12.68$	&$1.18$	\\
$0$	&$-2$	&$5$	&$12.53$	&$1.21$	\\
$0$	&$-3$	&$-$	&$11.33$	&$1.31$	\\
$0$	&$-3$	&$0.05$	&$12.12$	&$1.20$	\\
$0$	&$-3$	&$0.5$	&$13.82$	&$1.18$	\\
$0$	&$-3$	&$5$	&$12.05$	&$1.22$	\\
$300$	&$0$	&$-$	&$20.88$	&$0.99$	\\
$300$	&$0$	&$0.05$	&$14.53$	&$1.02$	\\
$300$	&$0$	&$0.5$	&$19.67$	&$0.91$	\\
$300$	&$0$	&$5$	&$16.52$	&$1.05$	\\
$300$	&$-1$	&$-$	&$11.65$	&$1.30$	\\
$300$	&$-1$	&$0.05$	&$12.65$	&$1.14$	\\
$300$	&$-1$	&$0.5$	&$18.27$	&$0.97$	\\
$300$	&$-1$	&$5$	&$16.39$	&$1.08$	\\
$300$	&$-2$	&$-$	&$8.23$	&$1.38$	\\
$300$	&$-2$	&$0.05$	&$15.75$	&$1.10$	\\
$300$	&$-2$	&$0.5$	&$17.01$	&$1.02$	\\
$300$	&$-2$	&$5$	&$12.73$	&$1.11$	\\
$300$	&$-3$	&$-$	&$11.58$	&$1.26$	\\
$300$	&$-3$	&$0.05$	&$15.45$	&$1.11$	\\
$300$	&$-3$	&$0.5$	&$16.35$	&$1.08$	\\
$300$	&$-3$	&$5$	&$15.08$	&$1.13$	\\
\hline\hline
\end{tabular}
\end{table}

\section{Discussion}
\label{sec:discussion}

In this paper, we model a single stellar population formed in an instantaneous burst at $t = 0$, considering different stellar rotation velocities ($v = 0$ and $300\,\mathrm{km,s^{-1}}$), metallicity ($[\ce{Fe/H}] = 0$, $-1$, $-2$, and $-3$), ISM densities ($n_{\mathrm{ISM}} = 0.05$, $0.5$, and $5\,\mathrm{cm^{-3}}$), and IMFs (Kroupa, log-flat, and Chon). We investigate the evolution of dust produced by core-collapse supernovae in the early Universe, explicitly accounting for the effects of the reverse shock.

The adopted dust yields and grain size distributions are taken from the recent work of \citet{otaki2026}, in which dust formation is modeled following a fixed-energy supernova explosion expanding into a uniform ISM.
We note that our theoretical models do not account for the effects of galaxy geometry or dust column density when computing the extinction curves. 

Nevertheless, we also compare our results with the extinction curves inferred for quasars at $z\sim 4\text{--}7$ by \citet{Gallerani2010}, which probe extinction rather than attenuation and therefore provide a more direct comparison with dust models. They found extinction curves flatter than the SMC, although still not as flat as some of the recent high-redshift galaxy attenuation curves.
This supports the idea that early dust populations can produce intrinsically shallow extinction curves, while the even flatter attenuation curves observed in galaxies likely require additional effects related to dust geometry, optical depth, and radiative transfer.

To investigate attenuation curves in greater detail and enable a more direct comparison with observations, we plan to explore these properties of early-Universe galaxies using cosmological simulations with the \texttt{dustyGadget} code \citep{graziani2020}. In these simulations, we will incorporate dust mass yields and self-consistent grain size distributions from core-collapse supernovae \citep{otaki2026}.

Our results can also be placed in the context of previous theoretical studies of SN dust extinction at high redshift. \citet{bianchi2007} modeled the surviving dust mass after accounting for reverse-shock destruction, while neglecting dust dynamics. They found that, compared to no-reverse-shock models, the extinction curve computed from IMF-averaged grain size distributions becomes steeper in the far-UV at wavelengths $\lesssim 2000\,\text{\AA}$, in good agreement with the extinction curve of a quasar at $z \sim 6$ observed by \citet{maiolino2004}.
Our results are consistent with their prediction of a broad bump feature at $\sim 2500\,\text{\AA}$ produced by AC grains. However, the far-UV rise attributed to \ce{Fe3O4} grains in their model is not present in most of our cases, because the surviving fraction of \ce{Fe3O4} after the passage of the reverse shock is generally low (see \citealt{otaki2026}).

Similarly, \citet{hirashita2008} derived theoretical extinction curves based on the grain size distributions of Pop III core-collapse supernovae, including reverse-shock destruction \citep{nozawa2007}. They concluded that models in high-density environments ($n_\mathrm{H}\gtrsim 1,\mathrm{cm^{-3}}$) are consistent with the observed extinction properties at $z\sim6$. In agreement with their findings, our models also predict flatter extinction curves as the ISM density increases.
Nevertheless, unlike our results, their extinction curves do not show a pronounced bump at $\sim 2500\,\text{\AA}$, since silicate grains dominate their dust composition. This difference likely arises from variations in the adopted pre-supernova progenitor models and the resulting dust species distributions.

Our results indicate that the depletion of small grains by the reverse shock produces relatively flat extinction curves with a broad bump at $2500\,\text{\AA}$, rather than the distinct $2175\,\text{\AA}$ feature.
However, on timescales of $\sim 100\,\mathrm{Myr}$, shattering of large grains with sizes $\gtrsim 100\,\mathrm{nm}$ can efficiently replenish the population of small grains ($\lesssim 10\,\mathrm{nm}$) through turbulence in the warm ionized medium \citep[][see also e.g., \citealt{narayanan2023}]{hirashita2010}. This mechanism could naturally explain the observed redshift evolution of extinction curves, which show an increasing bump strength toward lower redshift \citep{witstok2023b, markov2023, yang2023, Ormerod2025, lin2025}.
In addition, carbon grains ejected by supernovae that survive the destructive effects of the reverse shock may undergo further processing in the ISM, fragmenting into smaller particles and potentially forming PAHs that give rise to the $2175\,\text{\AA}$ bump \citep{yang2023}.

Future work will therefore include a more comprehensive treatment of dust processing within the ISM, going beyond the pure dust yields from supernova explosions.

\section{Conclusions}
\label{sec:conclusions}

The main results of this study can be summarized as follows:
\begin{itemize}
\item The extinction curves at $30 \,\mathrm{Myr}$ exhibit significant diversity, depending on the grain species and size distributions associated with the properties of SN progenitors, such as mass, rotation velocity, and metallicity. Most models do not show the steep slope characteristic of the SMC extinction curve. Instead, they display a broad bump at $\sim 2500,\text{\AA}$, produced by large AC grains. In addition, the reverse shock tends to flatten the extinction curves through the efficient destruction of small grains.

\item Compared to the Kroupa IMF, the top-heavy IMF proposed by Chon et al. (2022) yields higher total dust masses at $30 \,\mathrm{Myr}$ in all models. In particular, the nonrotating models produce steeper extinction curves with larger $A_\lambda/A_V$, whereas the rotating models tend to yield flatter curves. In the nonrotating case, fragmentation of large grains enhances the contribution of smaller grains, leading to steeper slopes. By contrast, rotating progenitors form larger grains prior to the passage of the reverse shock; consequently, the survival of large grains and the depletion of small grains result in flatter extinction curves.

\item In the no-reverse-shock cases, the contribution of small grains generally increases between $7 \,\mathrm{Myr}$ and $30 \,\mathrm{Myr}$, producing progressively steeper extinction curves over time. In contrast, when small grains are destroyed by reverse shocks, the extinction curve shows little temporal evolution, except in the nonrotating model with $[\ce{Fe/H}] = -2$. In this case, the contribution of grains with sizes in the range $10$–$100\,\mathrm{nm}$ increases with time, leading to a steeper extinction curve at $30 \,\mathrm{Myr}$.
\end{itemize}

Overall, our results demonstrate that the destructive effects of SN reverse shocks can significantly modify extinction curves in the early Universe. This study provides insight into how extinction curves evolve in galaxies hosting young stellar populations, as a result of dust production and processing by core-collapse supernovae.

Our findings contribute to ongoing efforts to incorporate physically motivated dust models into cosmological simulations and to improve our understanding of the evolution of attenuation curves in high-redshift galaxies recently observed by JWST \citep[e.g.,][]{markov2023, witstok2023b, Markov2025, shivaei2025}.

\section{Data availability}
The IMF-averaged grain size distributions and normalized extinction curves at $7$, $10$, and $30\,\mathrm{Myr}$ for all models presented in this work are available at \url{https://github.com/K-Otaki/dust-enrichment-and-extinction-curves}. All data related to this study are also available from the corresponding author upon reasonable request.

\begin{acknowledgements}
KO, RS, and LG acknowledge support from the PRIN 2022 MUR project 2022CB3PJ3-First Light And Galaxy aSsembly (FLAGS) funded by the European Union–Next Generation EU, from EU-Recovery Fund PNRR - National Centre for HPC, Big Data and Quantum Computing, and from the MUR project "Fondo Italiano per la Scienza" FIS-2024-01621 DAWN. ML acknowledges partial financial contribution from the PRIN 2022 (20224MNC5A) - "Life, death and after-death of massive stars" funded by European Union–Next Generation EU and from the INAF Theory Grant "Supernova remnants as probes for the structure and mass-loss history of the progenitor systems".
JW gratefully acknowledges support from the Cosmic Dawn Center through the DAWN Fellowship. The Cosmic Dawn Center (DAWN) is funded by the Danish National Research Foundation under grant No. 140.
This work is partially supported by MEXT as “Project for Establishment of a Center for Advanced HPC-AI Development Support”.
\end{acknowledgements}

%
\bibliographystyle{aa} 
\bibliography{refs} 
%
\begin{appendix}
\label{sec:appendix}

\section{Time-dependent properties}





\begin{figure}[htbp]
    \centering
    \includegraphics[width=\linewidth]{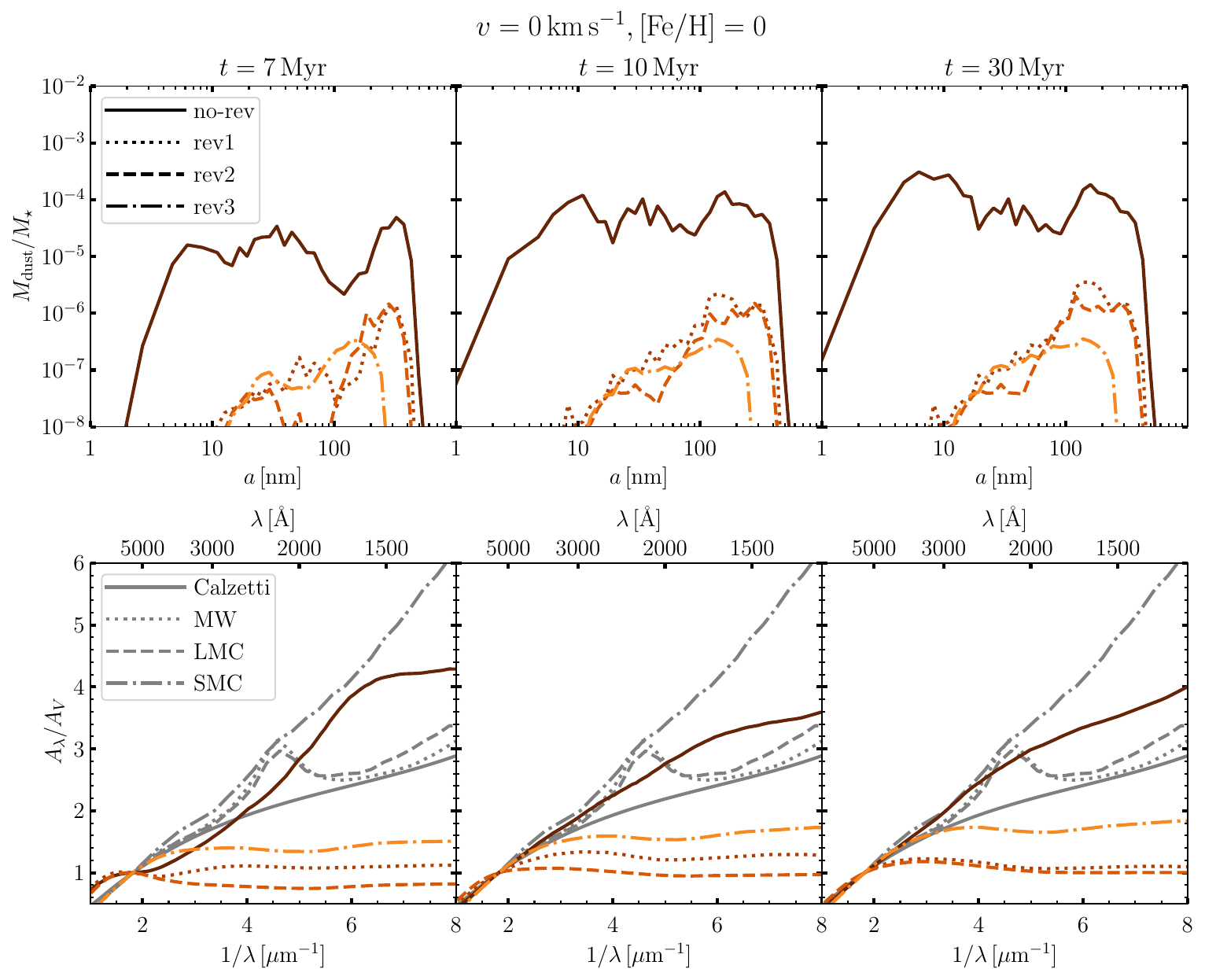}
    \caption{Time evolution of the grain size distributions (upper panels) and extinction curves (lower panels). From left to right, the panels show the results at $7$, $10$, and $30\,\mathrm{Myr}$ after the formation of nonrotating stars with $[\ce{Fe/H}] = 0$ in an instantaneous burst with a Kroupa IMF.}
    \label{fig:ext_distri_V0_Kroupa_Z1}
\end{figure}

\begin{figure}[htbp]
    \centering
    \includegraphics[width=\linewidth]{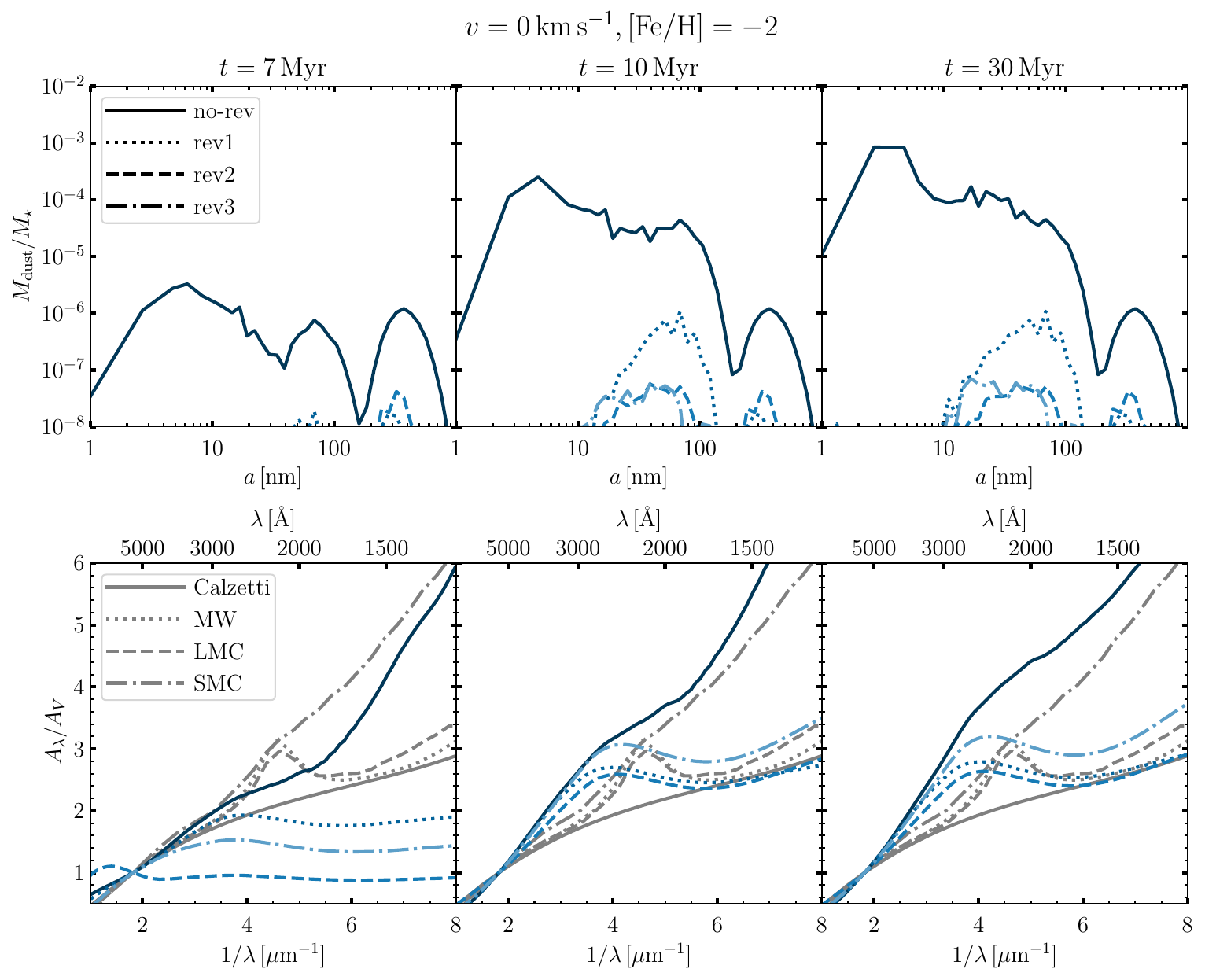}
    \caption{Same as Fig. \ref{fig:ext_distri_V0_Kroupa_Z1}, but for progenitors with $[\ce{Fe/H}]=-2$. }
    \label{fig:ext_distri_V0_Kroupa_Z3}
\end{figure}

\begin{figure}[htbp]
    \centering
    \includegraphics[width=\linewidth]{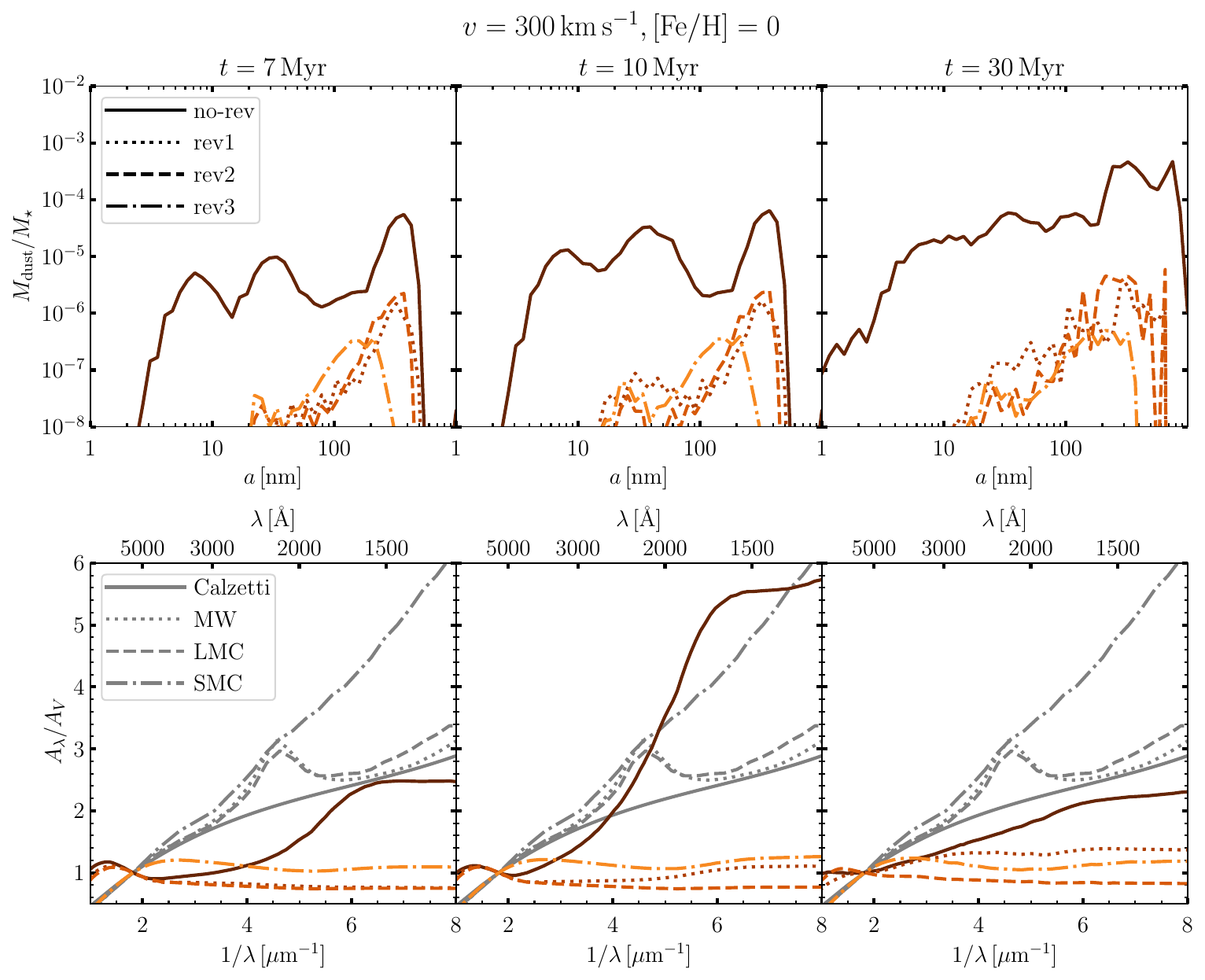}
    \caption{Time evolution of the grain size distributions (upper panels) and extinction curves (lower panels). From left to right, the panels show the results at $7$, $10$, and $30\,\mathrm{Myr}$ after the formation of rotating stars with $[\ce{Fe/H}] = 0$ in an instantaneous burst with a Kroupa IMF.
}
    \label{fig:ext_distri_V300_Kroupa_Z1}
\end{figure}

\begin{figure}[htbp]
    \centering
    \includegraphics[width=\linewidth]{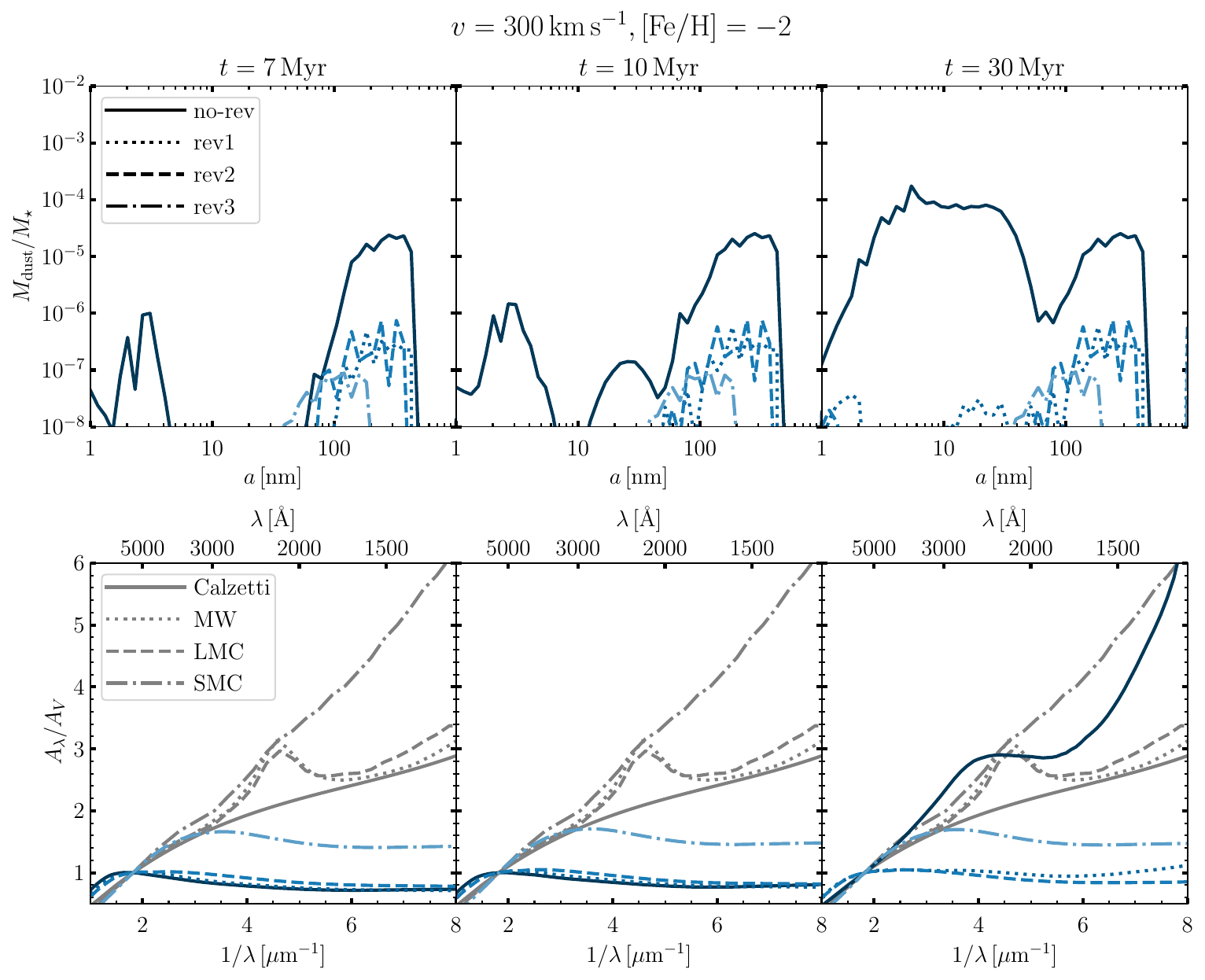}
    \caption{Same as Fig. \ref{fig:ext_distri_V300_Kroupa_Z1}, but for progenitors with $[\ce{Fe/H}]=-2$. }
    \label{fig:ext_distri_V300_Kroupa_Z3}
\end{figure}

In this Appendix, we present the time-dependent evolution of the predicted grain size distributions and extinction curves. 
To visualize how the grain size distributions and extinction curves change with time, we illustrate the results at three different times after the burst: 
$t = 7$, $10$, and $30\,\mathrm{Myr}$ (see Figures~\ref{fig:ext_distri_V0_Kroupa_Z1}--\ref{fig:ext_distri_V300_Kroupa_Z3}). The timescale of $7\,\mathrm{Myr}$ corresponds to the lifetimes of nonrotating and rotating progenitors with stellar masses of $m_\star \simeq 25\,M_\odot$ and $30\,M_\odot$, respectively. Consequently, the dominant contributors to dust production on this timescale are massive progenitor stars with masses $\gtrsim 25$--$30\,M_\odot$.


In general, the mass of small grains with sizes $\lesssim 100\,\mathrm{nm}$ increases with time due to the contribution of SN ejecta from progressively less massive progenitors. At the same time, the reverse shock reduces the abundance of very small grains ($\lesssim 10\,\mathrm{nm}$), limiting their survival as the system evolves.

For the nonrotating and rotating models with $[\ce{Fe/H}] = 0$, shown in Figures~\ref{fig:ext_distri_V0_Kroupa_Z1} and~\ref{fig:ext_distri_V300_Kroupa_Z1}, respectively, the extinction curves do not vary significantly over time and remain nearly flat despite the evolution of the grain size distributions. This behavior indicates that the relative abundances of small and large grains increase at comparable rates, without building up a sufficient population of very small grains to produce a pronounced UV bump.

In contrast, the nonrotating model with $[\ce{Fe/H}] = -2$ (Figure~\ref{fig:ext_distri_V0_Kroupa_Z3}) shows a stronger time dependence. At $7\,\mathrm{Myr}$, the extinction curve of the no-rev model resembles the SMC curve, with a steep UV slope, while the reverse-shock models exhibit flatter profiles. By $10\,\mathrm{Myr}$, all models develop a broad bump at $\sim 4\,\mathrm{\mu m^{-1}}$ and a steeper UV rise at $\gtrsim 6\,\mathrm{\mu m^{-1}}$. At $30\,\mathrm{Myr}$, the extinction curve of the no-rev model becomes similar to that of the rev3 model, despite their different grain size evolution histories.

For the rotating progenitor model with $[\ce{Fe/H}] = -2$ (Figure~\ref{fig:ext_distri_V300_Kroupa_Z3}), the extinction curves of the reverse-shock models remain nearly flat, with $A_\lambda/A_V \sim 1$ and little temporal evolution. In contrast, the no-rev model develops an extinction curve similar to that of the LMC by $30\,\mathrm{Myr}$.

\begin{figure}[htbp]
    \centering
    \includegraphics[width=\linewidth]{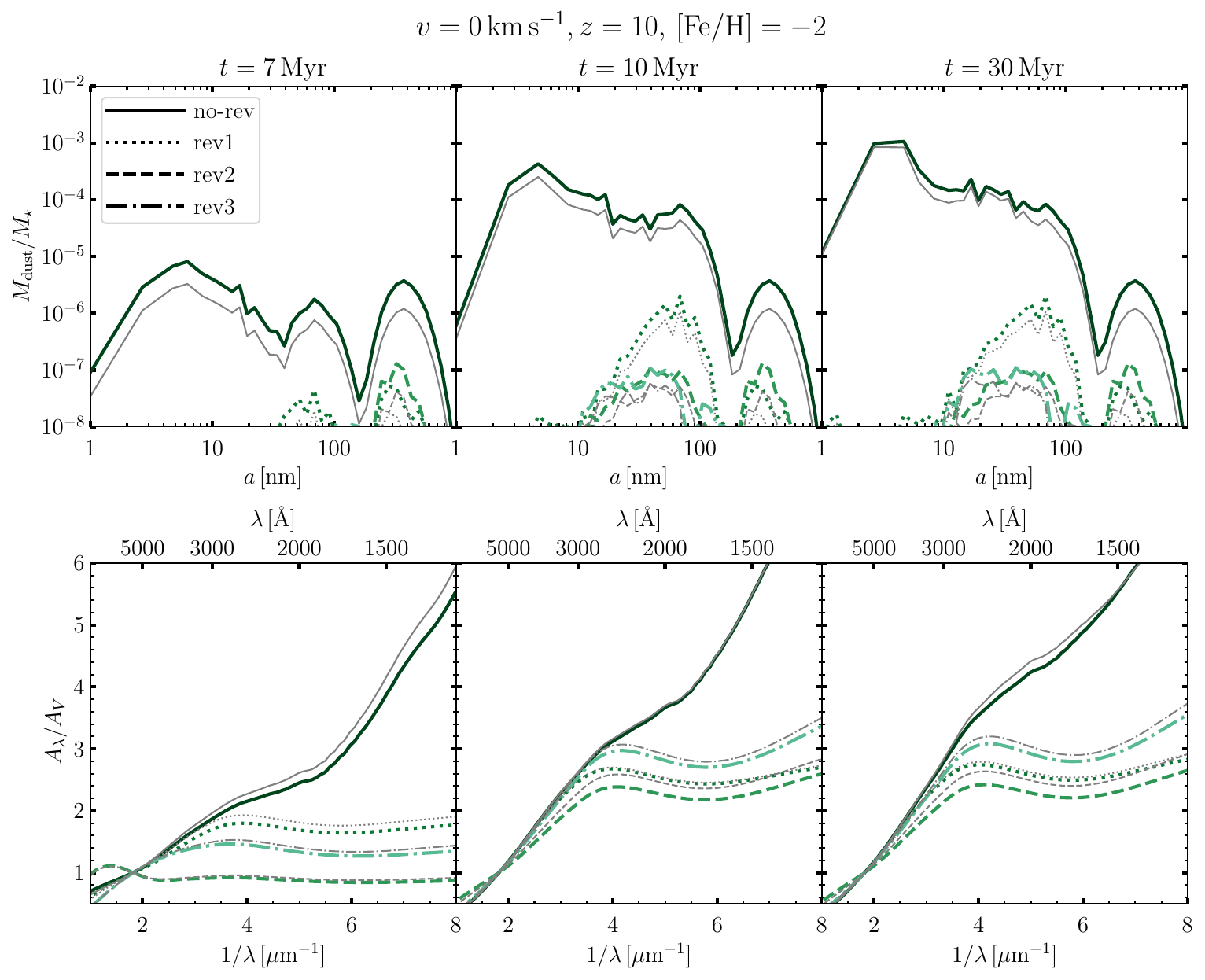}
    \caption{Time evolution of the grain size distributions (upper panels) and extinction curves (lower panels). From left to right, the panels show the results at $7$, $10$, and $30\,\mathrm{Myr}$ after the formation of nonrotating progenitor stars with $[\ce{Fe/H}] = -1$ in an instantaneous burst with the Chon IMF at $z=10$ (colored lines) and the Kroupa IMF (gray lines). In each panel, the solid, dotted, dashed, and dash-dotted lines correspond to the no-reverse-shock (no-rev), rev1, rev2, and rev3 models, respectively.
}
    \label{fig:comp_ext_distri_V0_Chon_Z2}
\end{figure}

\begin{figure}[htbp]
    \centering
    \includegraphics[width=\linewidth]{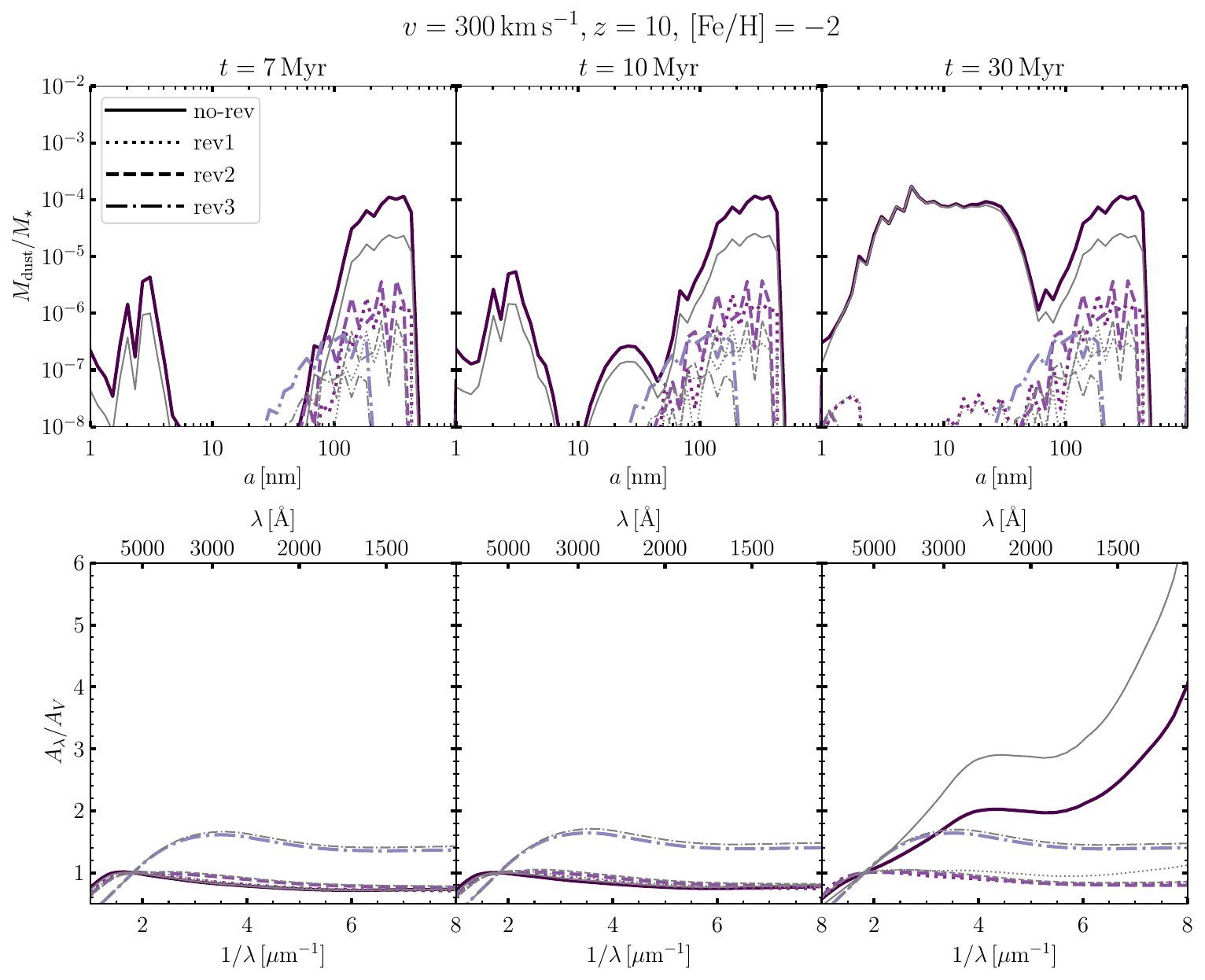}
    \caption{Same as Fig. \ref{fig:comp_ext_distri_V0_Chon_Z2}, but for rotating progenitor stars.}
    \label{fig:comp_ext_distri_V300_Chon_Z2}
\end{figure}

Figures~\ref{fig:comp_ext_distri_V0_Chon_Z2} and~\ref{fig:comp_ext_distri_V300_Chon_Z2} present the time evolution of the grain size distributions and extinction curves assuming the Chon IMF at $z = 10$, for nonrotating and rotating progenitors with $[\ce{Fe/H}] = -2$, respectively. In each panel, the colored lines correspond to the Chon IMF, while the gray lines represent the Kroupa IMF.

Adopting the top-heavy Chon IMF increases the fraction of massive stars and therefore enhances the total dust mass at all grain sizes compared to the Kroupa IMF. As a consequence of the modified size distribution, the extinction curves become slightly flatter, independently of the progenitor model. In particular, for the no-rev rotating model with $[\ce{Fe/H}] = -2$, the extinction curve obtained with the Chon IMF is flatter than that with the Kroupa IMF, owing to the larger fraction of grains with sizes $\gtrsim 100,\mathrm{nm}$ produced by massive progenitors.

At all epochs considered here, rotating models consistently predict flatter extinction curves than nonrotating models.

\end{appendix}

\end{document}